\documentclass[%
 reprint,
 amsmath,amssymb,
 aps,
]{revtex4-2}

\usepackage{graphicx}% Include figure files
\usepackage{dcolumn}% Align table columns on decimal point
\usepackage{bm}% bold math
\usepackage{xcolor}
\usepackage{epstopdf}
\usepackage{amsmath}
\usepackage{enumerate}
\begin{document}

\preprint{APS/123-QED}

\title{Dual CFT on Nariai limit for Kerr-Sen-dS black holes}% Force line breaks with \\
%\thanks{A footnote to the article title}%

\author{Muhammad Fitrah Alfian Rangga Sakti}
\email{fitrahalfian@gmail.com}
\author{Piyabut Burikham}%
\email{piyabut@gmail.com}
% \altaffiliation[Also at ]{Physics Department, XYZ University.}%Lines break automatically or can be forced with \\
\affiliation{High Energy Physics Theory Group, Department of Physics, Faculty of Science, Chulalongkorn University, Bangkok 10330, Thailand}

%
%\collaboration{MUSO Collaboration}%\noaffiliation
%
%\author{Charlie Author}
% \homepage{http://www.Second.institution.edu/~Charlie.Author}
%\affiliation{
% Second institution and/or address\\
% This line break forced% with \\
%}%
%\affiliation{
% Third institution, the second for Charlie Author
%}%
%\author{Delta Author}
%\affiliation{%
% Authors' institution and/or address\\
% This line break forced with \textbackslash\textbackslash
%}%
%
%\collaboration{CLEO Collaboration}%\noaffiliation

\date{\today}% It is always \today, today,
             %  but any date may be explicitly specified

\begin{abstract}
In this work, we study the Kerr-Sen-de Sitter black hole~(BH) in the Nariai limit where the event and cosmological horizon coincide. We show that the near-horizon Kerr-Sen-de Sitter black hole in Nariai limit is a fiber over AdS$_2$ with an appropriate coordinate transformation, instead of fiber over dS$_2$. Hence, we can compute the associated central charge and CFT temperature by using the Kerr/CFT method. It is remarkably exhibited that through Cardy's growth of states, the Bekenstein-Hawking entropy on cosmological horizon is reproduced. Moreover, we show that the radial equation of the quantum scalar field in $J$- and $Q$-pictures on this charged rotating background in Nariai limit can be portrayed in quadratic Casimir operator form with $SL(2,R)\times SL(2,R)$ isometry. We also compute the corresponding thermodynamic quantities from CFT to find the absorption cross-section and real-time correlator in $J$-picture. In $Q$-picture, we do not find a well-defined CFT description. We then extend the study of quantum scalar field in Nariai limit for Kerr-Newman-dS black hole solution and show that the hidden conformal symmetry on this black hole's background in in $J$- and $Q$-pictures is well-defined.
%\begin{description}
%%\item[Usage]
%%Secondary publications and information retrieval purposes.
%\item[PACS numbers]
%04.40.Nr, 04.60.-m, 04.70.Dy
%%\item[Structure]
%%You may use the \texttt{description} environment to structure your abstract;
%%use the optional argument of the \verb+\item+ command to give the category of each item. 
%\end{description}
\end{abstract}

%\pacs{04.20.Jb, 04.50.Kd, 04.70.Dy}% PACS, the Physics and Astronomy
                             % Classification Scheme.
%\keywords{Suggested keywords}%Use showkeys class option if keyword
                              %display desired
\maketitle

%\tableofcontents

\section{Introduction}
\label{sec:intro}
Establishing the holographic correspondence between de Sitter space (dS) and quantum theory is a significant challenge in theoretical physics. The holographic duality, most notably captured by the Anti-de Sitter/Conformal Field Theory (AdS/CFT) correspondence, has provided profound insights into the quantum theory of gravity. However, extending this framework to dS space has proven to be more problematic. One of the main difficulties in the holographic description of dS space arises from the absence of a global timelike Killing vector and spatial infinity. In the AdS/CFT correspondence, the asymptotic behavior of AdS space allows for a well-defined boundary at infinity, which plays a crucial role in the holographic interpretation. This boundary serves as a holographic screen, encoding the dynamics of the gravitational theory in terms of a dual quantum field theory living on the boundary. In contrast, dS space does not possess a similar asymptotic structure or boundary at infinity. As a result, the standard techniques used in the AdS/CFT correspondence cannot be straightforwardly applied to dS space. It is an intriguing and challenging problem that requires a deeper understanding of the quantum nature of gravity, as well as the dynamics of dS space itself.

Inspite of the challenges coming from the asymptotically de Sitter spacetime, this solution may provide a correct interpretation to explain our physical world in the context of cosmological model. The standard model of cosmology, whose details are explained
extensively e.g. in \cite{MukhanovCambridgePress2006}, is surprisingly consistent with this de Sitter solution. The observed accelerated expansion of our universe, supported by astronomical observations such as the cosmic microwave background radiation and the distribution of galaxies, suggests that our universe could be described by a dS spacetime \cite{PeeblesRatraRevModPhys2003,PadmanabhanPhysRep2003}. In particular, the connection between the cosmological constant and the behavior of smaller objects, such as black holes, is also of great interest. The presence of a non-zero cosmological constant affects the configurations and properties of black holes. In fact, the cosmological constant can influence the formation, growth, and evaporation of black holes. Moreover, one can study the quantum gravity from black hole's properties through the gravitational waves observations \cite{AlexanderFinnYunesPRD2008,AbediDykaarPRD2017,OshitaTsunaPRD2020}. Hence, in the future, we expect that the gravitational wave observations may give us more precise information about quantum structure of black holes including when the cosmological constant is present.

Regarding the black hole solutions, one of the central challenges in quantum gravity is to understand the microscopic origin of black hole entropy, which is given by the famous Bekenstein-Hawking formula stating that the entropy is one quarter of the black hole's event horizon area. Remarkably, the AdS/CFT correspondence provides a potential framework for addressing this problem. It is assumed that the partition function of conformal field theory to be identical to black hole's partition function. This relation was firstly proven by computing the entropy the extremal Kerr black hole \cite{GuicaPRD2009}. That study is inspired from the success of the investigation of the asymptotic symmetries of Banados-Teitelboim-Zanelli (BTZ) black hole which contains two-dimensional (2D) local conformal algebra.  CFT techniques can then be used to calculate the state degeneracy \cite{StromingerJHEP1998} leading to the agreement of CFT entropy with the Bekenstein-Hawking entropy.

In the following, we consider a class of charged rotating black hole solution in Einstein-Maxwell-Dilaton-Axion (EMDA) supergravity theory and Einstein-Maxwell theory, namely dyonic Kerr-Sen-dS (KSdS) and dyonic Kerr-Newman-dS (KNdS) black holes, respectively. Both black holes possess more than two horizons because of the existence of the positive cosmological constant. The largest horizon is the cosmological horizon for which the first thermodynamics law on this horizon resembles the first thermodynamic law on black hole's horizon. In the cosmological horizon, one can also study the thermal spectrum with non-zero entropy that is also proportional to the horizon area, likewise in the event horizon. The existence of three horizons leads to different limits. When the Cauchy and event horizons coincide, one can obtain an extremal black hole. When the cosmological and event horizon coincide, the resulting metric will be Nariai solution. Furthermore, when all three horizons coincide, the ultracold solution will be produced.

The first holographic description of the cosmological entropy of the Nariai solution obtained from Kerr-dS and KNdS solutions has been investigated in Ref.~\cite{Anninos2010}. The Nariai solution possesses $SL(2,R)\times U(1)$ isometry where the spacetime metric is a fiber over dS$_2$. We want to show that in Nariai limit, we can write the solution with similar isometry, although with the spacetime metric as a fiber over AdS$_2$ by choosing different sign on the radial near-horizon coordinate. This metric form on Nariai limit denotes the similar form with near-horizon extremal metric \cite{Sakti2018,SaktiEPJPlus2019,SaktiAnnPhys2020,SaktiPhysDarkU2021}. Furthermore, we want to study the scalar wave equation on the geometry when we take the Nariai limit. As has been studied for generic Kerr solution \cite{Castro2010}, we exhibit that in Nariai limit, the scalar wave equation possesses $SL(2,R)\times SL(2,R)$ isometry. This isometry happens to appear in scalar wave equation of some black hole solutions \cite{Saktideformed2019,SaktiNucPhysB2020,SaktiPhysDarkU2022,Saktidyonicnonex} and also for higher-spin field \cite{ChenLong2010,ChenLongJHEP2010,DjogamaSaktiarXiv2023}. Remarkably, one can also compute the absorption cross-section and real-time correlator from 2D CFT of the solution on Nariai limit. We want to prove that  for both KNdS and KSdS black holes in Nariai limit, the conformal symmetry can be shown on the spacetime metric directly and on the scalar wave equation. Moreover, CFT dual will be constructed trough the matching between the entropy, absorption cross-section, and the real-time correlator. We will consider two different pictures, $J$- and $Q$ pictures. $J$- picture denotes the CFT description when the scalar probe is neutral while $Q$-picture denotes the CFT description when the scalar probe is electrically charged.

We organize the paper as follows. In Sec. \ref{sec:KSdS}, we carry out the KSdS solution and its thermodynamic quantities on the event and cosmological horizons. In Sec. \ref{sec:SS}, we show the "fin" diagrams which portray the horizons and the extremal limits. In the next section, we study the Nariai limit on KSdS solution and find the entropy using the Kerr/CFT correspondece. In Sec. \ref{sec:scatteringonKSdS}, we study the conformal symmetry on the scalar wave equation in $J$- and $Q$-pictures in the background of KSdS solution in Nariai limit. In Sec. \ref{sec:KNdS}, we further extend the study of the scalar field on KNdS black hole in Nariai limit. In Sec. \ref{sec:summary}, we summarize our findings for the whole paper.

\section{Dyonic Kerr-Sen-de Sitter solution and its thermodynamics }\label{sec:KSdS}

The dyonic KSdS black hole is the exact solution to EMDA supergravity theory with positive cosmological constant. The line element of KSdS spacetime reads as \cite{WuWuWuYuPRD2021}
\begin{equation}
	ds^2 = - \frac{\Delta}{\varrho^2 } \hat {X}^2+  \frac{\varrho ^2}{\Delta}d\hat{r}^2 + \frac{\varrho ^2}{\Delta_\theta} d\theta^2 + \frac{\Delta_\theta \sin^2\theta}{\varrho ^2}\hat{Y}^2,\label{eq:KSdSmetric} \
\end{equation}
where
\begin{equation}
	\hat{X} = d\hat{t} - a \sin^2\theta \frac{d\hat{\phi}}{\Xi},~\hat{Y} = ad\hat{t}- (\hat{r}^2-d^2-k^2+a^2 ) \frac{d\hat{\phi}}{\Xi},\nonumber\
\end{equation}
%\vspace{-0.7cm}
\begin{equation}
	\Delta = (\hat{r}^2-d^2-k^2+a^2)\left(1-\frac{\hat{r}^2-d^2 -k^2}{l^2} \right)-2m\hat{r}+p^2 +q^2, \nonumber\ \label{DelAdSshifted}\
\end{equation}
%\vspace{-0.7cm}
\begin{equation}
	\Delta_\theta = 1+\frac{a^2}{l^2}\cos^2\theta, ~\Xi =1+\frac{a^2}{l^2},~
	\varrho^2 =\hat{r}^2-d^2 -k^2 + a^2\cos^2\theta.\
\end{equation}
Note that $m,a,d,k,p, q, l$ are the parameters of mass, spin, dilaton charge, axion charge, magnetic charge, electric charge, and dS length, respectively. The KSdS solution above is obtained from the gauged dyonic Kerr-Sen black hole solution \cite{WuWuWuYuPRD2021,SaktidyonicKerrSen} via analytic continuation \cite{DolanKastorPRD2013}. There is an important relation between charges of KSdS solution given by
\begin{equation}
	d= \frac{p^2 -q^2}{2m}, ~~~~~ k =\frac{pq}{m}.\
\end{equation}
Therefore, when the magnetic charge vanishes, the axion charge also vanishes while when $p=q$, the dilaton charge will vanish. One also can write
\begin{equation}
	d^2+k^2 = \left( \frac{p^2+q^2}{2m}\right)^{2}. \label{eq:relationdkpq}\
\end{equation}

KSdS black hole possesses three positive horizons. The three positive horizons are inner ($r_-$), outer/event ($r_+$) and cosmological ($r_c$) horizons where $r_c\geq r_+\geq r_-\geq 0 $. We can study the thermodynamics on its horizons. Within this paper, we emphasize the study on thermodynamics on the event horizon and cosmological horizon. On the event horizon, the KSdS black hole satisfies the thermodynamic relation
\begin{equation}
	dM= T_H dS_{BH}+\Omega_H dJ +\Phi_H dQ +\Psi_H dP + Vd\mathcal{P}, \label{eq:thermoKSdS}
\end{equation}
with the following quantities
\begin{equation}
	M= \frac{m}{\Xi}, ~~~ J = \frac{ma}{\Xi}, ~~~ Q= \frac{q}{\Xi}, ~~~ P= \frac{p}{\Xi},\label{eq:MJQPKSdS}\
\end{equation}
%\vspace{-0.2cm}
\begin{equation}
	T_H = \frac{r_+(l^2-2r_+^2+2d^2 +2k^2-a^2)-ml^2}{2\pi (r_+^2 -d^2-k^2 +a^2)l^2}, \label{eq:THKSdS}
\end{equation}
%\vspace{-0.5cm}
\begin{equation}
	S_{BH}=\frac{\pi}{\Xi}(r_+^2 -d^2 -k^2 +a^2), ~~~
	\Omega_H = \frac{a\Xi}{r_+^2 -d^2 -k^2 +a^2}, \label{eq:SOmegaKSdS}
\end{equation}
%\vspace{-0.5cm}
\begin{equation}
	\Phi_H = \frac{q(r_+ +d -p^2/m)}{r_+^2 -d^2 -k^2 +a^2},~	\Psi_H = \frac{p(r_+ +d -p^2/m)}{r_+^2 -d^2 -k^2 +a^2}, \label{eq:PhiPsiKSdS}
\end{equation}
%\vspace{-0.5cm}
\begin{equation}
	V= \frac{4}{3}r_+ S_{BH}, ~~~\mathcal{P}=-\frac{3}{8\pi l^2}. \label{eq:VdKSdS}
\end{equation}
where those are physical mass, angular momentum, electric charge, magnetic charge, Hawking temperature, Bekenstein-Hawking entropy, angular velocity, electric potential, magnetic potential, volume and pressure. We can also consider the thermodynamic quantities on the cosmological horizon $r_c$. Those thermodynamic quantities on $r_c$ are given as follows
\begin{equation}
	M_c=-\frac{m}{\Xi}, ~~~ J_c =-\frac{ma}{\Xi}, ~~~ Q_c=-\frac{q}{\Xi}, ~~~ P_c=-\frac{p}{\Xi},\label{eq:MJQPKSdScos}\
\end{equation}
%\vspace{-0.2cm}
\begin{equation}
	T_c = \frac{r_c(2r_c^2-2d^2-2k^2+a^2-l^2)+ml^2}{2\pi (r_c^2 -d^2-k^2 +a^2)l^2}, \label{eq:THKSdScos}
\end{equation}
%\vspace{-0.5cm}
\begin{equation}
	S_{c}=\frac{\pi}{\Xi}(r_c^2 -d^2 -k^2 +a^2), ~~~
	\Omega_c = \frac{a\Xi}{r_c^2 -d^2 -k^2 +a^2}, \label{eq:SOmegaKSdScos}
\end{equation}
%\vspace{-0.5cm}
\begin{equation}
	\Phi_c = \frac{q(r_c +d -p^2/m)}{r_c^2 -d^2 -k^2 +a^2},~	\Psi_c = \frac{p(r_c +d -p^2/m)}{r_c^2 -d^2 -k^2 +a^2}, \label{eq:PhiPsiKSdScos}
\end{equation}
%\vspace{-0.5cm}
\begin{equation}
	V_c= \frac{4}{3}r_c S_c, ~~~\mathcal{P}_c=\frac{3}{8\pi l^2}. \label{eq:VdKSdScos}
\end{equation}
These thermodynamic quantities are obtained by considering the black hole's event horizon as the boundary \cite{GomberoffPRD2003}. This is in contrast with the previous thermodynamic quantities on the black hole's event horizon where the cosmological horizon is considered as the boundary. For this black hole, we consider the cosmological constant as dynamical variable as a consequence of considering mass of the black hole as enthalpy of the spacetime \cite{CveticGibbonsPRD2011,DolanKastorPRD2013}.

\section{Structure of Dyonic Kerr-Sen-de Sitter Spacetime}
\label{sec:SS}
\begin{figure*}
\begin{tabular}{cc}
	\includegraphics[width=0.5\linewidth]{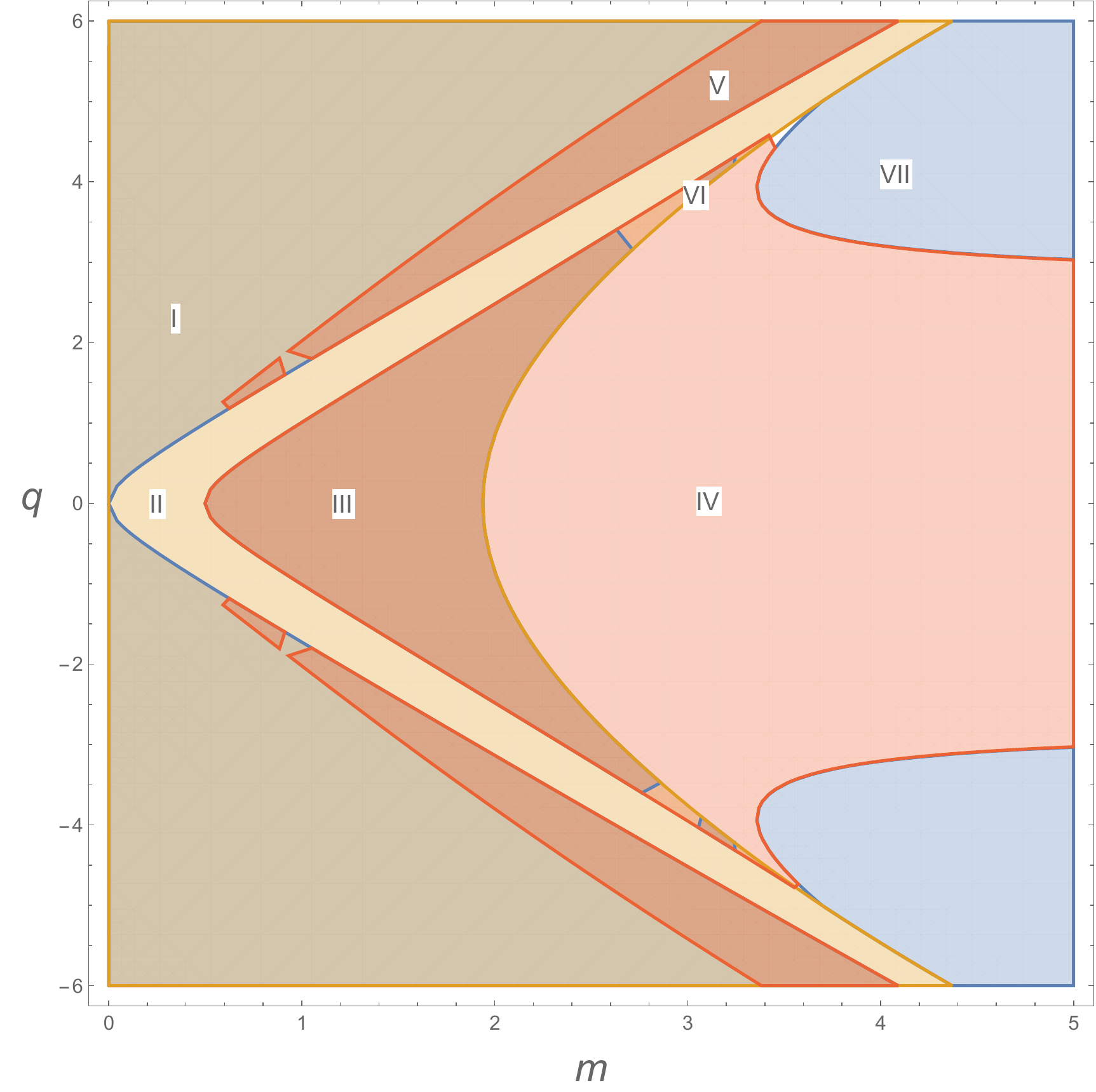}&
	\includegraphics[width=0.5\linewidth]{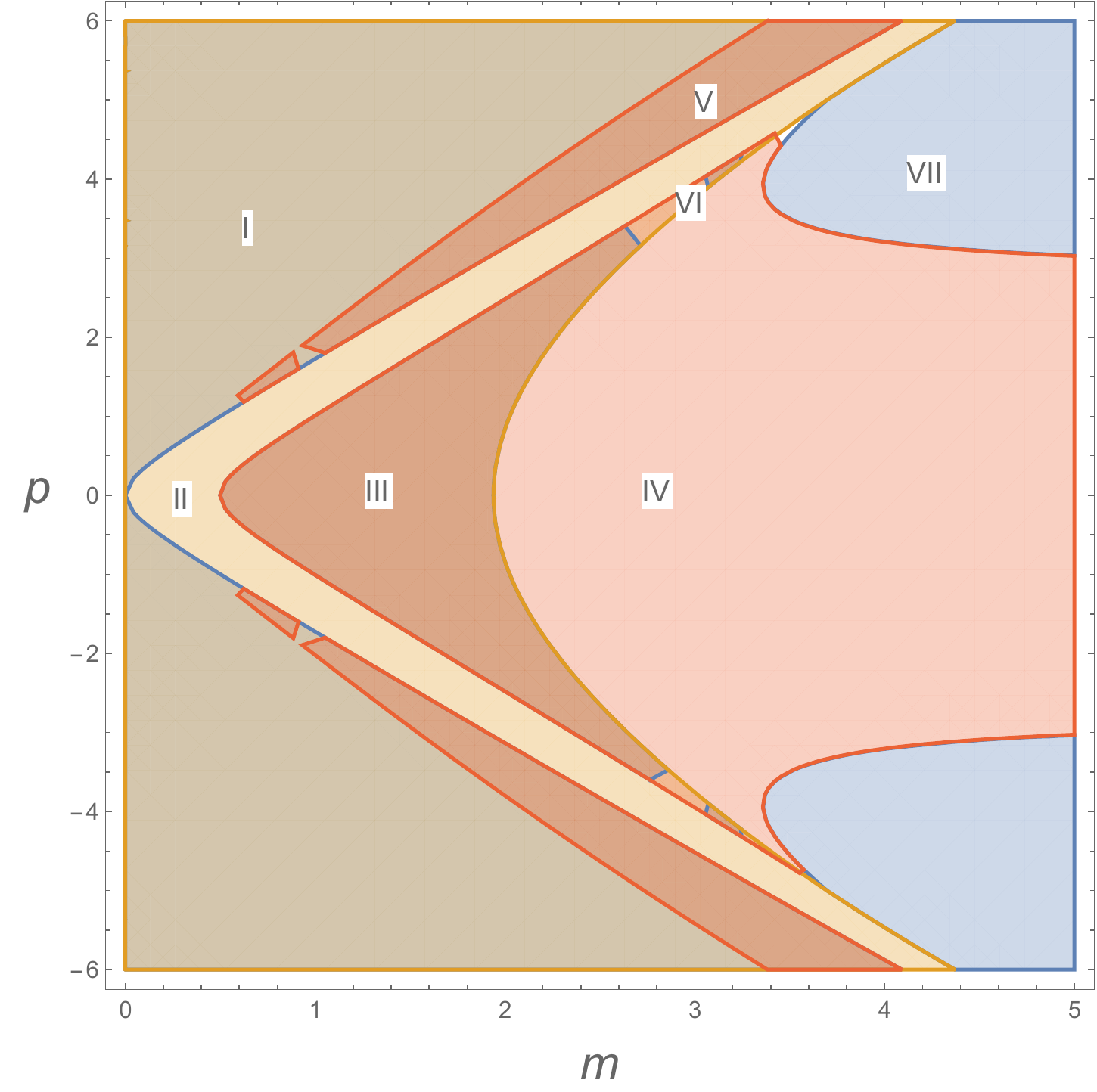} \\
(a) $p=0$ KSdS& (b) $q=0$ KSdS\\
	\includegraphics[width=0.5\linewidth]{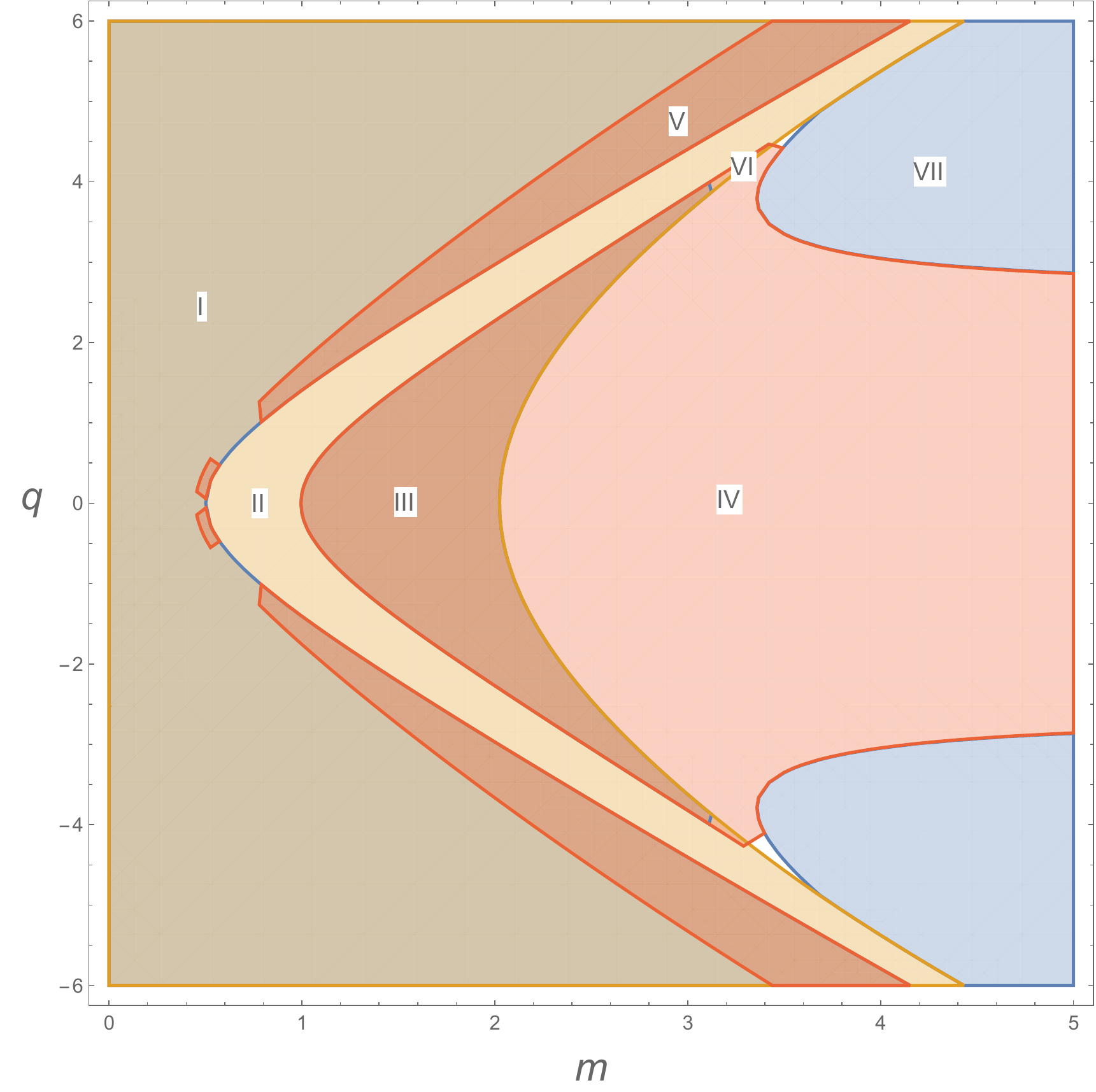}&
      \includegraphics[width=0.5\linewidth]{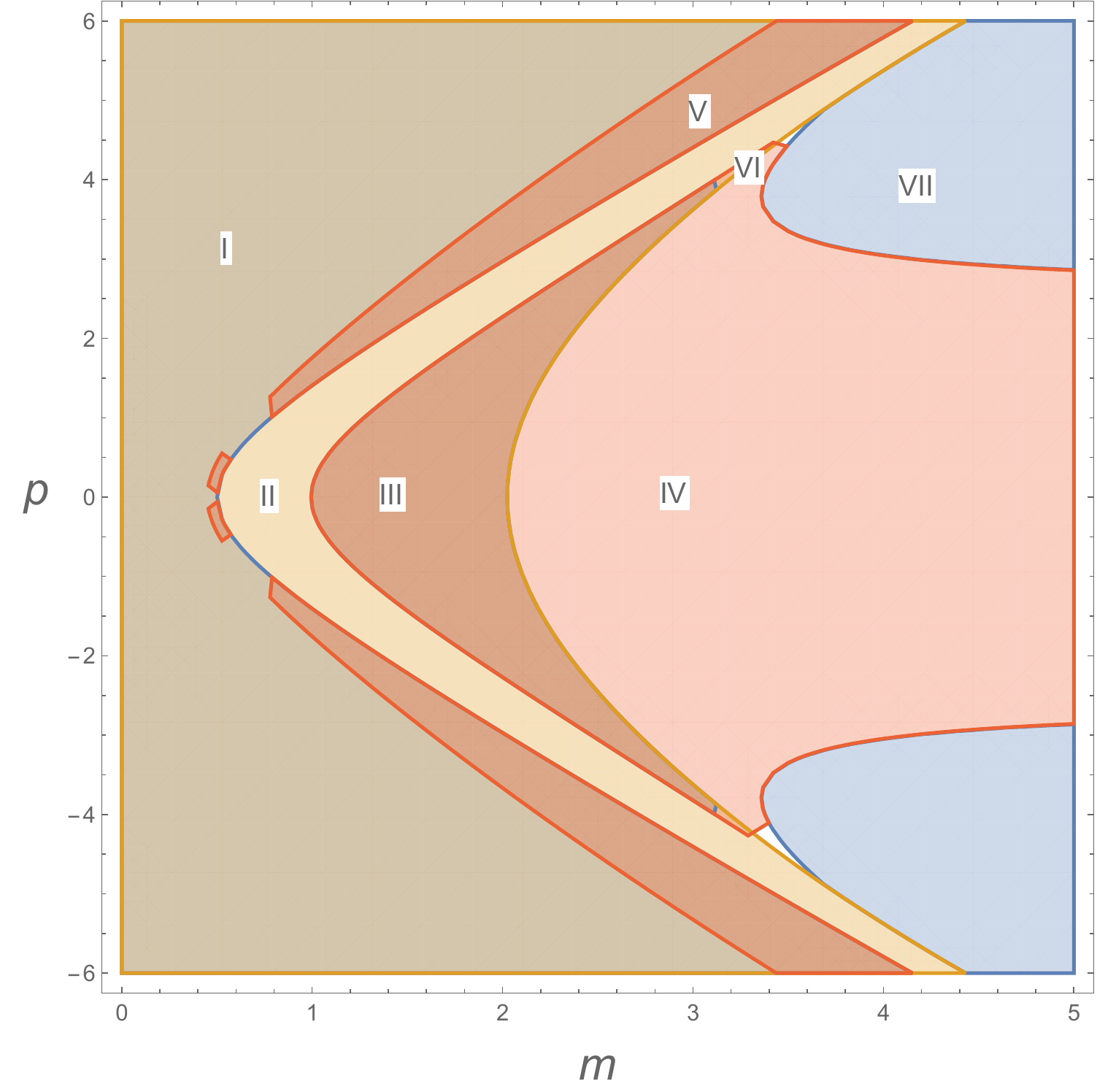} \\
(c) $p=1$ KSdS & (d) $q=1$ KSdS \\	
\end{tabular}
\caption{Horizon structure of KSdS spacetime for $a=0.5, l = 10$, color region represents parameter space with the horizon $r_{i}\geq 0$ for $i=1-4$.  Region I consists of an outer horizon behind a naked singularity and cosmological horizon. Region II, IV, and VII are spacetime with naked singularity and cosmological horizon. Region III and VI are black hole spacetime with singularity behind the Cauchy horizon, event horizon, and cosmological horizon. Region V contains naked singularity and cosmological horizon, and inner and outer horizon of black hole locating behind a naked singularity. }\label{figFin}
\end{figure*}

In order to gain insight into the structure of dyonic KSdS spacetime, we explore the horizon solutions by plotting the ``fin" diagram by slicing through the parameter space with 4 planes; (a) $p=0$, (b) $q=0$, (c) $p=1$, (d) $q=1$, for $a=0.5, l = 10$ as shown in Fig.~\ref{figFin}. The zeroes of metric function $\Delta/\varrho^2$ give four roots for the horizons, $\hat{r}=r_{i},~(i=1-4)$. We observe the continuity of the real part of horizon ${\rm Re}[r_i]$ connecting solutions of merging horizons. In addition, there is a ring singularity at $\varrho = 0$ or $\hat{r}=\sqrt{d^{2}+k^{2}}=\displaystyle{\frac{p^{2}+q^{2}}{2m}}\equiv r_{s}$.

For zero-charge spinning KSdS spacetime at fixed $a, l$, there is an extremal horizon where both inner~(Cauchy) horizon $r_{-}$ and outer~(black hole) horizon $r_{+}$ emerge at the critical $m$, above which $r_{-}$ and $r_{+}$ move away from one another. Finally at certain mass, $r_{+}$ will merge with cosmological horizon $r_{c}$ and the outer physical spacetime region disappears leaving the naked singularity at $\hat{r}=0$.  

When either electric~($q$) or magnetic~($p$) charge is turned on, the spacetime structure becomes more complicated as depicted in Fig.~\ref{figFin}. Region I contains an outer horizon, naked singularity and cosmological horizon where $r_{+}<r_{s}<r_{c}$. Region II, IV, and VII are spacetime with naked singularity and cosmological horizon and no black holes. Region III and VI are spacetime with black holes and cosmological horizon, there exists singulairty hidden behind the Cauchy horizon. Region V intriguingly has black hole region with inner and outer horizon behind the naked singularity and cosmological horizon at the furthest. 

The boundary lines between each Region where extremal horizons occur are categorized as follows.

\begin{enumerate}[(i)]
\item I and II: $r_{-}=r_{+}<r_{s}$,
\item II and III+VI: $r_{-}=r_{+}>r_{s}$,
\item III+VI and IV: $r_{+}=r_{c}>r_{s}$, Nariai BH,
\item II and V: $r_{-}=r_{+}<r_{s}$, 
\item I and V: $r_{-}=0, r_{+}<r_{s}$, 
\item IV and VII: $r_{1}=r_{4}\equiv r_{c}>r_{s}$, 
\item II and IV, VII: $r_{1}=r_{2}\equiv r_{c}>r_{s}$.
\end{enumerate}

The Nariai limit of all cases occurs in the high mass region where $r_{+}=r_c$, the boundary line between region III+VI and IV. On the other hand, the boundary line between region II and III+VI represents extremal solutions where $r_{-}=r_{+}$. In comparison to the ``shark fin" figure in Ref.~\cite{Castro:2022cuo}, the dependence of $d, k$ on $m$ and  $a$ distort the fin shape in the small mass and charge region in KSdS case. Interestingly, similar kinds of ``flag" diagram where extremal structures are explored for generalized spherically symmetric metric are also presented in Ref.~\cite{Burikham:2020dfi,Wuthicharn:2019olp}.

\par
\section{Charged Rotating Nariai/CFT Correspondence}
\label{sec:NariaiCFTKSdS}
The rotating Nariai solutions from the Kerr-dS and KNdS solutions haves been obtained in Ref. \cite{Anninos2010}. In this section, we will exhibit that this rotating solution in Nariai limit can also be expressed in the metric form with a fiber over AdS$_2$ as the common near-horizon metric form in the Kerr/CFT correspondence \cite{Sakti2018,SaktiEPJPlus2019,SaktiAnnPhys2020,SaktiPhysDarkU2021}. Another purpose of this section is to give a more detailed derivation of the central charge, temperature and entropy for solution in Nariai limit obtained from KSdS black hole. Furthermore, we will also study the massless scalar wave equation on this rotating background in Nariai limit which may possesses $SL(2,R)\times SL(2,R)$ isometry.

\subsection{Geometry of Nariai Limit}
To find the rotating geometry in Nariai limit, we will assume two different limits which are Nariai limit and near-horizon limit. The parameter that parameterizes the Nariai limit is defined by
\begin{equation}
\varepsilon =\frac{r_c - r_+}{\lambda r_0}.\label{eq:Nariaipar}
\end{equation}
$r_0$ is a scaling constant that we define as $r_0^2=r_+^2-d^2-k^2+a^2$ for KSdS solution. When $\varepsilon \rightarrow 0$, we can obtain $r_c =r_+$. Furthermore, the near-horizon coordinate transformations are given by \cite{Hartman2009}
\begin{equation}
\hat{r}=r_+ +\lambda r_0 r, ~~ \hat{t}=\frac{r_0}{\lambda}t, ~~\hat{\phi}=\phi+\frac{\Omega_c r_0}{\lambda}t.\label{eq:nearhorizontrans}\
\end{equation}
%\begin{equation}
%	\hat{r}=r_c+\lambda r_0 r, ~~ \hat{t}=\frac{r_0}{\lambda}t, ~~\hat{\phi}=\phi+\frac{\Omega_c r_0}{\lambda}t.\label{eq:nearhorizontrans}\
%\end{equation}
The constant $\lambda$ is to parameterize the near-horizon limit. $\lambda\rightarrow 0$ denotes the near-horizon limit. Using those coordinate transformations, one can find the charged rotating solution in Nariai limit as follows
%\begin{eqnarray}
%ds^2 &=& \Gamma(\theta)\left(-r(\epsilon+r) dt^2 + \frac{dr^2}{r(\epsilon+r)} + \alpha(\theta) d\theta ^2 \right)  \nonumber\\
%&& +\gamma(\theta) \left(d\phi +e r dt\right)^2, \label{eq:rotatingNariaimetric}\
%\end{eqnarray}
\begin{eqnarray}
	ds^2 &=& \Gamma(\theta)\left(-r(r-\epsilon) dt^2 + \frac{dr^2}{r(r-\epsilon)} + \alpha(\theta) d\theta ^2 \right)  \nonumber\\
	&& +\gamma(\theta) \left(d\phi +e r dt\right)^2, \label{eq:rotatingNariaimetric}\
\end{eqnarray}
where the metric functions are given by
%\begin{equation}
%\Gamma(\theta)= \frac{\varrho_c ^2}{\upsilon} , ~~\alpha(\theta) = \frac{\upsilon}{\Delta_\theta}, ~~\gamma(\theta)=\frac{r_0^4 \Delta_\theta \sin^2\theta}{\varrho_c^2 \Xi^2},\nonumber\
%\end{equation}
\begin{equation}
	\Gamma(\theta)= \frac{\varrho_+ ^2}{\upsilon} , ~~\alpha(\theta) = \frac{\upsilon}{\Delta_\theta}, ~~\gamma(\theta)=\frac{r_0^4 \Delta_\theta \sin^2\theta}{\varrho_+^2 \Xi^2},\nonumber\
\end{equation}
\begin{equation}
		\varrho_+^2 = r_+^2 -d^2-k^2+ a^2\cos^2\theta, ~~
	e = \frac{2ar_+\Xi}{r_0^2 \upsilon}. \label{eq:constant_e}
\end{equation}
%\begin{equation}
%\varrho_c^2 = r_c^2 + a^2\cos^2\theta, ~~
%e = \frac{2ar_c\Xi}{r_0^2 \upsilon}. \label{eq:constant_e}
%\end{equation}
Note that we already approximate the function $\Delta$ as
\begin{equation}
\Delta \simeq \upsilon (r-r_c)(r-r_+), \label{eq:deltapprox}\
\end{equation}
where $\upsilon = 1-(6r_+^2-d^2-k^2+a^2)/l^2$. The rotating solution in Nariai limit (\ref{eq:rotatingNariaimetric}) is a fiber over AdS$_2$. This fact can obviously be seen from the factor on time-like and radial coordinates of the metric in Nariai limit. One can also write the spacetime metric (\ref{eq:rotatingNariaimetric}) in Poincar{\' e} coordinates by applying the following coordinate transformations \cite{Compere2017}
%\begin{eqnarray}
%t &=&\frac{2}{\varepsilon}\text{log}\frac{y}{\sqrt{\tau^2y^2-1}},\\
%r&=&-\frac{\varepsilon}{2}(1+\tau y),\\
%\phi&=&\varphi+\frac{1}{2}\text{log}\frac{\tau y+1}{\tau y-1}.\
%\end{eqnarray}
\begin{eqnarray}
	t &=&-\frac{2}{\varepsilon}\text{log}\frac{y}{\sqrt{\tau^2y^2-1}},\\
	r&=&\frac{\varepsilon}{2}(1+\tau y),\\
	\phi&=&\varphi+\frac{1}{2}\text{log}\frac{\tau y+1}{\tau y-1}.\
\end{eqnarray}
From those transformations, one can find 
\begin{eqnarray}
ds^2 &=& \Gamma(\theta)\left(-y^2 d\tau^2 + \frac{dy^2}{y^2} + \alpha(\theta) d\theta ^2 \right)  \nonumber\\
&&+\gamma(\theta) \left(d\varphi +e y d\tau\right)^2, \label{eq:rotatingNariaimetricPoincare}\
\end{eqnarray}
This metric form is obviously similar with the common near-horizon metric that is used in the Kerr/CFT correspondence. This metric possesses $SL(2,R)\times U(1)$ isometry generated by the following vector fields
\begin{equation}
	\zeta_0 = \partial_\varphi , \label{eq:rotationalU1}\
\end{equation}
which denote the rotational $U(1)$ isometry and
\begin{equation}
	X_1 = \partial_\tau, ~~~ X_2 = \tau\partial_\tau - y \partial_y, \nonumber\
\end{equation}
\vspace{-0.6cm}
\begin{equation}
X_3 = \left(\frac{1}{2y^2}+\frac{\tau^2}{2} \right)\partial_\tau -\tau y \partial_y - \frac{e}{y}\partial_\varphi ,\label{eq:isometrynearhorizon}
\end{equation}
denoting $SL(2,R)$ isometry. 

As we have mentioned in the beginning of this section, in Ref. \cite{Anninos2010} they manage to show another form of charged rotating Nariai geometry from dyonic KNdS solution that is a fiber over dS$_2$ in Eq. (B.2) in their appendix, which is given by
\begin{eqnarray}
	ds^2 &=& \Gamma(\theta)\left(-(1-r^2) dt^2 + \frac{dr^2}{1-r^2} + \alpha(\theta) d\theta ^2 \right)  \nonumber\\
	&& +\gamma(\theta) \left(d\phi +k r dt\right)^2. \label{eq:rotatingNariaimetric2}\
\end{eqnarray}
For the detail forms of $\Gamma(\theta),\alpha(\theta),\gamma(\theta), k$, one can see in the reference that we have mentioned. This metric can be found by using the near-horizon coordinate transformations (2.13) in Ref.~\cite{Anninos2010} then using the coordinate changes (2.18). In that paper, they use $\hat{r}=r_c -\lambda r_c r$ in Eq. (2.13), instead of using positive new radial coordinate, that will result in metric with a fiber over dS$_2$. This is different with our result which is a fiber over AdS$_2$. Hence, the charged rotating geometry in Nariai limit can be portrayed in both fibers over dS$_2$ and AdS$_2$, depending on the sign of the near-horizon radial coordinate transformation we choose. In the other words, we can also obtain the charged rotating solution of KSdS black hole in Nariai limit with a fiber over dS$_2$ when we use $\hat{r}=r_c -\lambda r_c r$.

\subsection{CFT Duals}
For rotating Nariai geometry in Ref. \cite{Anninos2010}, they basically apply the similar Kerr/CFT method to study the relation between black hole's thermodynamics macroscopically and microscopically using 2D CFT. They propose the similar fall-off conditions for the metric deviations although the spacetime structure contains dS$_2$ slice, unlike in the common dual CFT which contains AdS$_2$ slice.

\subsubsection{Central charge}
In the Kerr/CFT correspondence, the entropy is assumed to originate from partition function of 2D CFT resulting in Cardy entropy formula. The main ingredients for this formula are the central charges and temperatures. The basic procedure to compute the central charges is to employ the approach of Brown and Henneaux \cite{brown1986} where the asymptotic symmetry group (ASG) needs to satisfy some certain boundary conditions. In this work, we consider the similar boundary conditions for the metric deviations as given in Ref. \cite{GuicaPRD2009},
\begin{eqnarray}
	h_{\mu \nu} \sim \left(\begin{array}{cccc}
		\mathcal{O}(r^2) & \mathcal{O}\left(\frac{1}{r^2}\right) &  \mathcal{O}\left(\frac{1}{r}\right) &  \mathcal{O}(1) \\
		&  \mathcal{O}\left(\frac{1}{r^3}\right) &  \mathcal{O}\left(\frac{1}{r^2}\right) &  \mathcal{O}\left(\frac{1}{r}\right) \\
		&  & \mathcal{O}\left(\frac{1}{r}\right) &  \mathcal{O}\left(\frac{1}{r}\right)\\
		&  &  &  \mathcal{O}(1)\
	\end{array} \right).\label{eq:gdeviation}
\end{eqnarray}
in the basis $ (t,r,\theta, \phi) $ for metric (\ref{eq:rotatingNariaimetric}). These metric deviations are subleading with respect to the background metric. These boundary conditions are basically chosen to eliminate excitations above extremality. The asymptotic symmetry of the general black hole family includes diffeomorphisms $ \xi $ that satisfy
\begin{equation}
	\delta_\xi g_{\mu\nu} = \mathcal{L}_\xi g_{\mu\nu} =  \xi^\sigma (\partial_\sigma g_{\mu\nu}) + g_{\mu\sigma}(\partial _\nu \xi^\sigma)+ g_{\sigma \nu}(\partial _\mu \xi^\sigma), \ 
\end{equation}
where the metric deviation is denoted by $ \delta_\xi g_{\mu\nu} = h_{\mu\nu} $.

The most general diffeomorphism symmetry that preserves such boundary conditions ({\ref{eq:gdeviation}) in the asymptotic infinity is generated by the following Killing vector field
	\begin{eqnarray}
		\zeta &=& \left[c_t + \mathcal{O}\left(r^{-3}\right) \right]\partial_t + \left[-r\epsilon '(\phi) + \mathcal{O}(1) \right]\partial _r \nonumber\\
		&& + \mathcal{O}\left(r^{-1}\right)\partial _\theta  + \left[\epsilon(\phi) + \mathcal{O}\left(r^{-2}\right) \right]\partial _\phi ,\
	\end{eqnarray}
where $ c_t $ is an arbitrary constant and the prime $ (') $ denotes the derivative with respect to $ \phi $. This ASG contains one copy of the conformal group of the circle which is generated by
\begin{eqnarray}
\zeta_\epsilon = \epsilon(\phi)\partial_\phi - r\epsilon '(\phi)\partial_r ,\label{eq:killingASG}
\end{eqnarray}
that will be the part of the near-horizon metric in Nariai limit. We know that the azimuthal coordinate is periodic under the rotation $ \phi \sim \phi+2\pi $. Hence, we may define $ \epsilon_{\hat{n}} = -e^{-i\hat{n} \phi} $ and $ \zeta_\epsilon =\zeta_\epsilon(\epsilon_{\hat{n}} ) $. By the Lie bracket, the symmetry generator (\ref{eq:killingASG}) satisfies the Witt algebra,
\begin{eqnarray}
i[\zeta_{\hat{m}} , \zeta_{\hat{n}} ]_{LB} = ({\hat{m}} -{\hat{n}} )\zeta_{{\hat{m}} + {\hat{n}} }. \label{eq:Witt}\
\end{eqnarray}
${\hat{m}},{\hat{n}} $ are just integers. The zero mode is the azimuthal translation or $U(1)$ isometry, $\zeta_0=-\partial_\phi$.

The associated conserved charge is \cite{BarnichBrandt2002}
\begin{eqnarray}
	Q_{\xi} = \frac{1}{8\pi}\int_{\partial\Sigma} k^{g}_{\zeta}[h;g].
\end{eqnarray}
This given integral is over the boundary of a spatial slice. The contribution of the metric tensor on the central charge is given explicitly by
\begin{eqnarray}
k^{g}_{\zeta}[h;g] &=& -\frac{1}{4}\epsilon_{\rho\sigma\mu\nu} \bigg\{ \zeta ^{\nu} D^{\mu} h - \zeta ^{\nu} D_{\lambda} h^{\mu \lambda}  \nonumber\\
& & + \frac{h}{2} D^{\nu}\zeta^{\mu} - h^{\nu \lambda} D_{\lambda}\zeta^{\mu} + \zeta_{\lambda}D^{\nu}h^{\mu \lambda} \nonumber\\
& & \left. +\frac{h^{\lambda \nu}}{2}\left(D^\mu \zeta_\lambda + D_\lambda \zeta^\mu \right) \right\} dx^\rho \wedge dx^\sigma . \label{eq:kgrav} \
\end{eqnarray}
We should note that the last two terms in Eq. (\ref{eq:kgrav}) vanish for an exact Killing vector and an exact symmetry, respectively. The charge $ Q_{\zeta} $ generates symmetry through the Dirac brackets. The ASG possesses algebra which is given by the Dirac bracket algebra of the following charges \cite{BarnichBrandt2002}
\begin{eqnarray}
	\{ Q_{\zeta},Q_{\bar{\zeta}} \}_{DB} &=& \frac{1}{8\pi}\int k^{g}_{\zeta}\left[\mathcal{L}_{\bar{\zeta}}g;g \right] \nonumber\\
	&=& Q_{[\zeta,\bar{\zeta}]} + \frac{1}{8\pi}\int k^{g}_{\zeta}\left[\mathcal{L}_{\bar{\zeta}}\bar{g};\bar{g}\right] \label{eq:charges}.\
	\
\end{eqnarray}
By using (\ref{eq:killingASG}) and upon quantization, we can transform the Dirac bracket algebra into a commutation relation that allows us to interpret the classical central charge as a quantum central charge of the dual CFT. For the quantization, we replace the classical charges $Q_{\zeta}$ by their quantum counterpart $L_{{\hat{n}} }$,
\begin{eqnarray}
	Q_{\zeta} \equiv L_{{\hat{n}} } - x \delta_{{\hat{n}} ,0}, \label{eq:quantumQL}
\end{eqnarray}
so that we obtain the conserved charges algebra in quantum form, such that
\begin{eqnarray}
	\left[L_{\hat{m}} , L_{\hat{n}}  \right] = ({\hat{m}} - {\hat{n}} ) L_{{\hat{m}} +{\hat{n}} } + \frac{c_L}{12}\hat{m} ({\hat{m}} ^2-1)\delta_{{\hat{m}} + {\hat{n}} , 0}. 
\end{eqnarray}
$ x $ is a free parameter \textcolor{black}{to scale the central charge on the term $\sim \hat{m}$ (see Appendix \ref{app:Virasoro})}. The relation (\ref{eq:Witt}) and (\ref{eq:quantumQL}) have been used to find the Virasoro algebra above. For the charged rotating solution in Nariai limit obtained from KSdS solution, the corresponding central charge is then
\begin{eqnarray}
c_L&=&\frac{3e}{\Xi}\int^\pi_0d\theta\sqrt{\Gamma(\theta)\alpha(\theta)\gamma(\theta)}\nonumber\\
&=&\frac{12ar_+}{1-\frac{6r_+^2-d^2-k^2+a^2}{l^2}}.\label{eq:cLKSdScos}\
\end{eqnarray}
The main difference between this central charge and the Nariai solution from KNdS black hole \cite{Anninos2010} is the presence of dilaton and axion charges.

\subsubsection{Temperature}
In the Nariai limit where $r_c \rightarrow r_+$, the variation of the entropy can be expressed completely in terms of a variation of the angular momentum, electric charge, magnetic charge, and the pressure. It takes the following form
%\begin{eqnarray}
%	dS_c = \frac{dJ_c}{T_L}+\frac{dQ_c}{T_q}+\frac{dP_c}{T_p}+\frac{d\mathcal{P}_c}{T_\mathcal{P}} ,\
%\end{eqnarray}
\begin{eqnarray}
	dS_{BH}= \frac{dJ}{T_L}+\frac{dQ}{T_q}+\frac{dP}{T_p}+\frac{d\mathcal{P}}{T_\mathcal{P}} ,\
\end{eqnarray}
where $T_L,T_q, T_p, T_\mathcal{P}$ are the left-moving temperature, conjugate temperature of electric charge,  conjugate temperature of magnetic charge, and  conjugate temperature of pressure, respectively. Hence, we can derive that
%\begin{eqnarray}
%T_L = - \frac{\partial T_c/\partial r_c}{\partial \Omega_c / \partial r_c}\bigg|_{Nariai} =\frac{\upsilon(r_c^2+a^2)}{4\pi ar_c \Xi}. \label{eq:TLKNdS}\
%\end{eqnarray}
\begin{eqnarray}
	T_L = - \frac{\partial T_H/\partial r_+}{\partial \Omega_H / \partial r_+}\bigg|_{r_+=r_c} \frac{\upsilon(r_+^2-d^2-k^2+a^2)}{4\pi ar_+ \Xi}. \label{eq:TLKSdScos}\
\end{eqnarray}
Since we already take $r_c\rightarrow r_+$, so the right-moving temperature is equal to zero which is proportional to the Hawking temperature on the cosmological horizon. However, in the next section, we will see that $T_R \sim \varepsilon$ generally which is fairly small.

\subsubsection{Cardy Entropy}
We have calculated the central extension and the corresponding temperature which are the main ingredients in Cardy entropy. So, the main upshot of this section is to provide the derivation of Cardy entropy \cite{GuicaPRD2009} for the charged rotating solution in Nariai limit. Firstly, Cardy entropy formula is given as follows
\begin{equation}
	S_{CFT}=\frac{\pi^2}{3}\left(c_L T_L +c_R T_R\right).\label{eq:Cardyentropy}
\end{equation}
The right-moving part is exactly zero, so only left-moving part contributes. On the cosmological horizon, by using above formula, we find the following entropy for charged rotating solution in Nariai limit
\begin{equation}
S_{CFT}=\frac{\pi}{\Xi}(r_+^2-d^2-k^2+a^2)=S_{BH}=S_c.\label{eq:entropyCFTNariai}\
\end{equation}
This result exhibits that the Kerr/CFT correspondence can be applied for charged rotating solution in Nariai limit obtained from KSdS black hole which precisely possesses $SL(2,R)\times U(1)$ isometry like the near-horizon extremal black hole. The entropy on the event horizon is now similar with the entropy on the cosmological horizon since we take $r_c \rightarrow r_+$. So, for an observer in the visible region, the total entropy in Nariai limit is \cite{SouravEPJC2016}
\begin{equation}
	S_{tot}=S_c+S_{BH}=2S_c.\label{eq:totalentropy}\
\end{equation}
Note that this is the total entropy for charged rotating solution in Nariai limit with non-vanishing dyonic charge. So, when $p=0$, it is just the total entropy for electrically charged one. Furthermore, when all electromagnetic charges vanish, it will be the total entropy of rotating Nariai solution obtained from Kerr-dS solution \cite{Anninos2010}. For the entropy on the event horizon for extremal KSdS black hole, we calculate in Appendix \ref{app:entropyKSdS} where we also prove the result in Ref.~\cite{SaktidyonicKerrSen} on KSdS solution that is mentioned therein. In case of asymptotically AdS spacetime, it has been carried out in Ref.~\cite{SaktidyonicKerrSen}. For vanishing cosmological constant and $p$, this result recovers the result in Ref. ~\cite{Ghezelbash2009}.

\section{Scattering of Scalar Field and Hidden Conformal Symmetry}
\label{sec:scatteringonKSdS}
In this section, we will further study the scattering of massless scalar field on the background of charged rotating Nariai solution in $J$ (angular momentum)- and $Q$ (electric charge)-pictures. We will also compute the absorption cross-section as well as the real-time correlator on the corresponding background. Yet, first we will show that there are hidden conformal symmetries on this solution in Nariai limit. Lastly, we also consider $Q$-picture where the scalar probe is assumed to be electrically charged.

\subsection{Scalar wave equation}
To explore the hidden conformal symmetries in $J$-picture, we assume a massless neutral scalar field in the background of charged rotating solution in Nariai limit. The massless scalar wave equation for the scalar probe is given by
\begin{equation}
\nabla_{\alpha} \nabla^{\alpha}\Phi = 0\label{KG1}.
\end{equation}
We know that the charged rotating solution in Nariai limit is conserved under time-like and azimuthal translations. Hence, we can separate the coordinates in the scalar wave equation as 
\begin{equation}
\Phi(\hat{t}, \hat{r}, \theta, \hat{\phi}) = \mathrm{e}^{- i \omega \hat{t} + i n \hat{\phi}} R(\hat{r}) S(\theta)\label{phi-expand1},
\end{equation}
where $\omega$ and $n$ are the asymptotic energy and angular momentum of the scalar field. Plugging Eq. (\ref{phi-expand1}) into Eq. (\ref{KG1}), leads to two differential equations i.e., the angular $S(\theta)$ and radial $R(\hat{r})$ wave functions,
\begin{eqnarray}
&&	\left[\frac{1}{\sin\theta} \partial_\theta (\sin\theta \partial_\theta )-  \frac{n^2\Xi^2}{\Delta_\theta\sin^2\theta} \right] S(\theta) \nonumber\\
&&+\left[\frac{2an\omega\Xi- a^2\omega^2 \sin^2\theta}{\Delta_\theta} \right] S(\theta) =- K_l S(\theta)  \label{eq:angular}, \
\end{eqnarray}
\begin{equation} 
	\bigg[ \partial_{\hat{r}} (\Delta \partial_{\hat{r}}) + \frac{\left[ (\hat{r}^2 -d^2-k^2+a^2) \omega - an\Xi \right]^2}{\Delta} - K_{l} \bigg] R(\hat{r}) = 0\label{eq:radial}, \
\end{equation}
where $K_{l}$ is the separation constant. Note that we still use the generic KSdS metric (\ref{eq:KSdSmetric}). These equations are identical with the wave equations in the Kerr-dS solutions. We do not consider the backreaction of the scalar field in this case.

Firstly, we will study the radial equation (\ref{eq:radial}) at the near region, which is defined by $ \omega \hat{r} \ll 1 $. Furthermore, we assume that the frequency of the scalar field to be very small $ \omega M \ll 1 $. Consequently, we also impose $ \omega a \ll 1, \, \omega d \ll 1,  $ and $\omega k \ll 1$. On the other hand, we also have $\omega q \ll 1,  $ and $\omega p \ll 1$. Since we have approximated $\Delta $ as (\ref{eq:deltapprox}), we may set up the radial equation in a suitable form for exploring its hidden conformal symmetry. The radial equation (\ref{eq:radial}) reduces to 
\begin{eqnarray}
&& \partial_{\hat{r}} \left[(\hat{r}-r_c)(\hat{r}-r_+) \partial_{\hat{r}}\right]R(\hat{r}) + \frac{r_c - r_+}{\hat{r} - r_c} A  R(\hat{r}) = 0 , \label{radeq} \nonumber\\
&& +  \left[\frac{r_c - r_+}{\hat{r} - r_+} B + C \right] R(\hat{r})=0, \label{eq:nearregionradialeqEMDA}
\end{eqnarray} 
where
\begin{equation}
A= \frac{ \left[ (r_c^2 -d^2-k^2 + a^2) \omega - a n \Xi \right]^2}{\upsilon^2 (r_c - r_+)^2}, \nonumber\
\end{equation}
\begin{equation}
B=-\frac{\left[ (r_+^2 -d^2-k^2 + a^2) \omega - a n \Xi \right]^2}{\upsilon^2(r_c - r_+)^2}, ~~~ C= \frac{-K_l}{\upsilon} . \label{CEMDA}\nonumber\
\end{equation}
It is worth noting that the angular wave equation does not have $ SL(2,R)\times SL(2,R) $ isometry, yet $ SU(2)\times SU(2) $ isometry \cite{LoweSkanataPRD2014}. Hence, we will not consider to investigate further the angular part.

In order to reveal the hidden symmetries of Eq. (\ref{eq:nearregionradialeqEMDA}), we need to perform the following conformal coordinate transformations \cite{Castro2010,Compere2017}
\begin{eqnarray}
&& \omega^c = \sqrt{\frac{\hat{r}-r_c}{\hat{r}-r_+}}e^{2\pi T_R \hat{\phi} + 2 n_R \hat{t}}, \label{con1}\nonumber\\
&& \omega^+ = \sqrt{\frac{\hat{r}-r_c}{\hat{r}-r_+}}e^{2\pi T_L \hat{\phi} + 2 n_L \hat{t}}, \label{con2}\nonumber\\
&& \hat{y} = \sqrt{\frac{r_c -r_+}{\hat{r}-r_+}}e^{\pi (T_L +T_R) \hat{\phi} + (n_L + n_R) \hat{t}}. \label{conformalcoord}\
\end{eqnarray}
Then we may define three locally conformal operators in terms of the new conformal coordinates $\omega^c,\,\omega^+$ and $\hat{y}$ as
\begin{eqnarray}
&& H_1 = i \partial_c, \label{vec1}\nonumber\\
&& H_{-1} = i \left(\omega^{c2}\partial_c + \omega^{c}\hat{y}\partial_{\hat{y}} - \hat{y}^2 \partial_+ \right), \label{vec2}\nonumber\\
&& H_0 = i \left(\omega^{c}\partial_c + \frac{1}{2}\hat{y}\partial_{\hat{y}} \right), \label{vec3}
\end{eqnarray}
as well as 
\begin{eqnarray}
&& \bar{H}_1 = i \partial_+, \label{vvec1}\nonumber\\
&& \bar{H}_{-1} = i \left(\omega^{+2}\partial_+ + \omega^{+}\hat{y}\partial_{\hat{y}} - \hat{y}^2 \partial_c \right), \label{vvec2}\nonumber\\
&& \bar{H}_0 = i\left(\omega^{+}\partial_+ + \frac{1}{2}\hat{y}\partial_{\hat{y}} \right). \label{vvec3}
\end{eqnarray}
Note that we have used $\partial_c =\partial/\partial\omega^c, \partial_+ =\partial/\partial\omega^+$. The set of operators (\ref{vec1}) satisfies the $ SL(2,R) $ Lie algebra
\begin{eqnarray}
\left[H_0,H_{\pm 1} \right] = \mp iH_{\pm 1}, ~~~ \left[H_{-1},H_1 \right]=-2iH_0, 
\end{eqnarray}
while a similar $ SL(2,R) $ algebra exists for the set of operators (\ref{vvec1}). From every set of operators, we can construct the quadratic Casimir operator as given by
\begin{eqnarray}
\mathcal{H}^2 &=& \bar{\mathcal{H}}^2 = - H_0^2 + \frac{1}{2}(H_1 H_{-1} + H_{-1} H_{1}) \nonumber\\
&=& \frac{1}{4}(\hat{y}^2 \partial_{\hat{y}}^2 - y\partial_{\hat{y}})+\hat{y}^2\partial_c \partial_+. \label{quadraticCasimir}
\end{eqnarray}

We have found that the radial equation (\ref{radeq}) could be re-written in terms of the $ SL(2,R) $ quadratic Casimir operator as $ \mathcal{H}^2 R(r)=\bar{\mathcal{H}}^2 R(r)= -C R(r) $ where, in $J$-picture, we should identify the constants as
\begin{equation}
	n_L = -\frac{\upsilon}{2(r_c + r_+)},~~~ n_R =0, \label{nJEMDA}
\end{equation}
\begin{equation}
	T_L = \frac{\upsilon(r_c^2 + r_+^2-2d^2-2k^2+2a^2)}{4\pi a(r_c + r_+)\Xi},~~~ T_R =\frac{\upsilon(r_c - r_+)}{4\pi a\Xi}. \label{eq:tempCFTKSdS}
\end{equation}
$ T_{R,L} $ are identified as the CFT temperatures that emerge as a result of the spontaneously symmetry breaking of the partition function on $ SL(2,R)\times SL(2,R) $ theory to the partition function of $ U(1)\times U(1) $ CFT. One can see that the periodic identification of the azimuthal coordinate $ \hat{\phi} \sim \hat{\phi} +2\pi $ causes the $ SL(2,R)\times SL(2,R) $ symmetry to spontaneously break down to $ U(1)\times U(1) $ symmetry by temperatures $T_R,T_L$,
\begin{equation}
\omega^c \sim e^{4\pi^2 T_R\omega^c},~ \omega^+ \sim e^{4\pi^2 T_L\omega^+},~
\hat{y} \sim e^{2\pi^2 (T_L+T_L)\hat{y}}.
\end{equation}
This identification is generated by the $ SL(2,R)\times SL(2,R) $ group element, $e^{-i4\pi^2 T_R H_0-i4\pi^2 T_L\bar{H}_0}$.

After finding the temperatures (\ref{eq:tempCFTKSdS}), we can also compute conjugate charges $E_L,E_R$ for this charged rotating solution. These conjugate charges can be obtained from the entropy via \cite{Compere2017}
\begin{equation}
\delta S_{CFT} = \frac{\delta E_L}{T_L}+\frac{\delta E_R}{T_R}.\label{eq:variationCFTentropy}
\end{equation}
In order to compute the conjugate charges, we consider the first law of thermodynamics for the charged rotating solution in Nariai limit (\ref{eq:thermoKSdS}) with the quantities (\ref{eq:MJQPKSdScos})-(\ref{eq:VdKSdScos}). Since we consider the neutral scalar field, we have $\delta Q= \delta P=\delta\mathcal{P}=0$. Now we can find the conjugate charges via $\delta S_{BH}=\delta S_{CFT}$. Using the identification $\delta M$ as $\omega$ and $\delta J$ as $n$ yields to the  identification of $\delta E _{L,R}$ as $\omega_{L,R}$.  Hence, in $J$-picture, we obtain that
\begin{eqnarray}
\omega _{L} = \frac{r_c^2 + r_+^2 -2d^2-2k^2 +2a^2}{2a\Xi}\omega, ~\omega _{R} = \omega _{L}-n.  \label{eq:omegaCFTgaugedEMDA} \
\end{eqnarray}
So, we find the left and right frequencies of 2D CFT for the charged rotating solution in Nariai limit for KSdS black hole in $J$-picture. For the study of the scalar field in KSAdS black hole, one can see in Ref.~\cite{Saktidyonicnonex}.

\subsection{Absorption cross-section in $J$-picture}\label{scat}
In the previous section, we have derived one remarkable result of the dual CFT, i.e. the entropy. Another realization of this duality is the equivalence of the absorption cross-section of the scalar probe. In investigating the absorption cross-section, we need to consider the asymptotic region. Since we have approximated $\Delta$ in near-horizon region, this approximation in the asymptotic region will break down, except we consider the Nariai limit. This is similar in the near-extremal case of the scalar wave equation \cite{Saktideformed2019,SaktiNucPhysB2020,ChenLongJHEP2010,ChenChen2011,ChenGhezelbash2011}. Beside the Nariai limit on (\ref{eq:Nariaipar}), we consider the near-horizon coordinate transformations (\ref{eq:nearhorizontrans}). We also consider the scalar probe with frequencies around the superradiant bound 
\begin{equation}
	\omega = n\Omega_H +\hat{\omega}\frac{\lambda}{r_0},\label{eq:superbound}
\end{equation}
where $\Omega_H$ are given by (\ref{eq:SOmegaKSdS}). We can re-write the radial equation (\ref{radeq}) by
%\begin{equation}
%\left[\partial_r\left[r\left(r+\varepsilon\right)\right]\partial_r +\frac{A_s}{r}+\frac{B_s}{r+\varepsilon}+C_s\right]R(r) =0,\label{eq:radialequationnearext}
%\end{equation}
\begin{equation}
	\left(\partial_r\left[r\left(r-\varepsilon\right)\right]\partial_r +\frac{A_s}{r-\varepsilon}+\frac{B_s}{r}+C_s\right)R(r) =0,\label{eq:radialequationnearext}
\end{equation}
where
%\begin{eqnarray}
%A_s = \frac{\hat{\omega}^2}{\upsilon^2 \varepsilon}, ~~~ B_s = - \frac{(\hat{\omega}-2nr_c\varepsilon\Omega_c)^2}{\upsilon^2\varepsilon},~~~ C_s=C_s(\hat{\omega}), \nonumber\
%\end{eqnarray}
\begin{eqnarray}
	A_s = \frac{(\hat{\omega}+2nr_+\varepsilon\Omega_H)^2}{\upsilon^2\varepsilon}, ~~~ B_s = - \frac{\hat{\omega}^2}{\upsilon^2 \varepsilon},~~~ C_s=C_s(\hat{\omega}), \nonumber\
\end{eqnarray}
and $ C_s $ is the new separation constant that is dependent on the frequency $\hat{\omega}$ that can be obtained by solving the angular wave equation. We then apply the coordinate transformation $ z =(r-\varepsilon)/r $. In this new radial coordinate, the radial equation (\ref{eq:radialequationnearext}) becomes
\begin{equation}
\left[z(1-z)\partial_z^2 + (1-z)\partial_z +\frac{\hat{A_s}}{z}+\hat{B_s}+\frac{C_s}{1-z}\right]R(z) =0,\label{eq:radialequationnearext0}
\end{equation}
where
%\begin{eqnarray}
%\hat{A_s} = \frac{\hat{\omega}^2}{\upsilon^2\varepsilon ^2}, ~~~\hat{B_s} = -\frac{(\hat{\omega}-2nr_c\varepsilon\Omega_c)^2}{\upsilon^2\varepsilon^2}. \nonumber\
%\end{eqnarray}
\begin{eqnarray}
	\hat{A_s} = \frac{(\hat{\omega}+2nr_+\varepsilon\Omega_H)^2}{\upsilon^2\varepsilon^2}, ~~~\hat{B_s} = -\frac{\hat{\omega}^2}{\upsilon^2\varepsilon ^2}. \nonumber\
\end{eqnarray}
The ingoing solution to differential equation (\ref{eq:radialequationnearext0}) is also given by hypergeometric function
\begin{equation}
	R(z)= z^{-i\sqrt{\hat{A_s}}}(1-z)^{1+h}~_2F_1(a_s,b_s;c_s;z), \label{eq:ingoingsolgauged}\
\end{equation}
with the parameters 
$a_s=1+h -i(\sqrt{\hat{A}_s}+\sqrt{-\hat{B}_s})$, $b_s=1+h -i(\sqrt{\hat{A}_s}-\sqrt{-\hat{B}_S})$, and $c_s = 1-2i\sqrt{\hat{A}_s}$. For this superradiant case, the relation between $h$ and $C_s$ is given by
\begin{equation}
	h=\frac{1}{2}\left(-1+ \sqrt{1-4C_s} \right).\label{eq:h}
\end{equation}
For the asymptotic region of the radial coordinate $\hat{r}$ (or equivalently $r\gg\varepsilon$), where  $z \sim 1$, the solution (\ref{eq:ingoingsolgauged}) reduces to
\begin{eqnarray}
	R(y) \sim D_0 r^{h}+D_1 r^{-1-h}, \label{eq:asymptotsol0}\
\end{eqnarray}
where
\begin{equation}
	D_0 = \frac{\Gamma(c_s)\Gamma(1+2h)}{\Gamma(a_s)\Gamma(b_s)}, ~~~ D_1 = \frac{\Gamma(c_s)\Gamma(-1-2h)}{\Gamma(c_s-a_s)\Gamma(c_s -b_s)}.\label{eq:gammadefinition0}\
\end{equation}
The conformal weight of the scalar field is equal to $h+1$. 

For the coefficient (\ref{eq:gammadefinition0}), we find the absorption cross-section of the scalar fields as 
\begin{equation}
	P_{abs} \sim \left| D_0 \right|^{-2} \sim \sinh \left( {2\pi \hat{A}_s^{1/2} } \right){\left| {\Gamma \left(a_s  \right)}  {\Gamma \left(b_s \right)} \right|^2 }\label{eq:Pabsgauged}.
\end{equation}
To be more supporting the correspondence between the charged rotating solution in Nariai limit and 2D CFT, we show that the absorption cross-section for the scalar fields  (\ref{eq:Pabsgauged}) can be obtained from the absorption cross-section in a 2D CFT \cite{Castro2010}
\begin{eqnarray}
	P_{abs} &\sim& {T _L}^{2h_L - 1} {T _R}^{2h_R - 1} \sinh \left( {\frac{{\omega_L }}{{2{T _L} }} + \frac{{\omega _R }}{{2{T _R} }}} \right)\nonumber\\
	&\times&\left| {\Gamma \left( {h_L + i\frac{{\omega _L }}{{2\pi {T _L} }}} \right)} {\Gamma \left( {h_R + i\frac{{\omega _R }}{{2\pi {T _R} }}} \right)} \right|^2.\label{eq:PabsCFT}\
\end{eqnarray}
The agreement between (\ref{eq:Pabsgauged}) and (\ref{eq:PabsCFT}) can be shown when we choose proper left and right frequencies $\omega_L,\omega_R$. Since in the previous subsection we have found the CFT temperatures and frequencies, we can directly calculate the similar quantities for the frequency of the scalar field in the near-superradiant bound by neglecting the second-order and higher corrections from $\lambda$. Hence, in terms of Nariai limit parameter and in near-superradiant bound, we obtain
%\begin{equation}
%	T_L = \frac{\upsilon}{4\pi\Omega_c r_c}, ~~~T_R= \frac{\upsilon \lambda r_0}{4\pi a\Xi}\varepsilon. \label{eq:temperatureCFTnariai}
%\end{equation}
\begin{equation}
	T_L = \frac{\upsilon}{4\pi\Omega_H r_+}, ~~~T_R= \frac{\upsilon \lambda r_0}{4\pi a\Xi}\varepsilon. \label{eq:temperatureCFTnariai}
\end{equation}
When we take the Nariai limit $\varepsilon\rightarrow 0$ the right-moving temperature will vanish. Then we recover the result for $T_R$ and $T_L$ in the previous section. The frequencies are now
\begin{equation}
	\omega_L = n, ~~~ \omega_R = \frac{r_0}{a\Xi}\left(\hat{\omega}+nr_+\varepsilon\Omega_c \right),\label{eq:omegasuper}\
\end{equation}
The conformal weights in CFT are given by
\begin{equation}
h_L=h_R =h+1.\label{eq:conformalweightCFT}\
\end{equation}
This gives rise to another nontrivial evidence to support the Kerr/CFT correspondence for black holes in Nariai limit.

\subsection{Real-time correlator in $J$-picture}
Furthermore, one can also compute the real-time correlator. The asymptotic behaviors of the scalar field with ingoing boundary condition on the background of charged rotating solution in Nariai limit (\ref{eq:asymptotsol0}) indicate that two coefficients possess different roles where $D_1$ indicates the response and $D_0$ indicates the source. Hence, the two-point retarded correlator is simply \cite{ChenChuCorrelatorJHEP2010,ChenLongCorrelatorJHEP2010}
\begin{equation}
	G_R \sim \frac{D_1}{D_0} = \frac{\Gamma(-1-2h)}{\Gamma(1+2h)}\frac{\Gamma(a_s)\Gamma(b_s)}{\Gamma(c_s-a_s)\Gamma(c_s-b_s)}.\label{eq:retardedcorr}\
\end{equation}
From Eq. (\ref{eq:retardedcorr}), it is easy to check that
\begin{equation}
	G_R \sim \frac{\Gamma(h_L - i\frac{\omega_L}{2T_L})\Gamma(h_R - i\frac{\omega_R}{2T_R})}{\Gamma(1-h_L - i\frac{\omega_L}{2T_L})\Gamma(1-h_R - i\frac{\omega_R}{2T_R})}.\label{eq:retardedcorr1}
\end{equation}
Then by using the relation $\Gamma(z)\Gamma(1-z)=\pi/\sin(\pi z)$, we can write above two-point retarded correlator into
\begin{eqnarray}
G_R &\sim& \sin\left(\pi h_L+i\frac{\omega_L}{2T_L}\right)\sin\left(\pi h_R+i\frac{\omega_R}{2T_R}\right)\nonumber\\
&&\bigg|\Gamma\left(h_L + i\frac{\omega_L}{2T_L}\right)\Gamma\left(h_R + i\frac{\omega_R}{2T_R}\right)\bigg|^2. \label{eq:correlator}
\end{eqnarray}
Since $h_L,h_R$ are integers, so we have
\begin{eqnarray}
&&\sin\left(\pi h_L+i\frac{\omega_L}{2T_L}\right)\sin\left(\pi h_R+i\frac{\omega_R}{2T_R}\right)\nonumber\\
&&=(-1)^{h_L+h_R}\sin\left(i\frac{\omega_L}{2T_L}\right)\sin\left(i\frac{\omega_R}{2T_R}\right). \label{eq:sinsorrelator}\
\end{eqnarray}

In CFT, the Euclidean correlator is given by
\begin{eqnarray}
&&G_E (\omega_{EL},\omega_{ER}) \sim T_L^{2h_L-1} T_R^{2h_R-1}e^{i\omega_{EL}/2T_L}e^{i\omega_{ER}/2T_R}\nonumber\\
&&\times \bigg|\Gamma\left(h_L + \frac{\omega_{EL}}{2T_L}\right)\Gamma\left(h_R + \frac{\omega_{EL}}{2T_R}\right)\bigg|^2.
\end{eqnarray}
Where we can define the Euclidean frequencies $\omega_{EL}=i\omega_L$, and $\omega_{ER}=i\omega_R$. It is important noting that $G_E$ corresponds to the values of retarded correlator $G_R$. The retarded Green function $G_R$ is analytic on the upper half complex $\omega_{L,R}$-plane. The value of $G_R$ along the positive imaginary $\omega_{L,R}$-axis gives the following correlator
\begin{equation}
	G_E(\omega_{EL},\omega_{ER}) = G_R(i\omega_{L},i\omega_{R}), ~~~ \omega_{EL,ER}>0. \label{eq:correlatormatch}
\end{equation}
At finite temperature, $\omega_{EL,ER}$ should take discrete values of the Matsubara frequencies, given by
\begin{equation}
	\omega_{EL} =2\pi m_L T_L, ~~~ \omega_{ER}=2\pi m_R T_R, \label{eq:Matsubara}
\end{equation}
where $m_L$ and $m_R$ are half integers for fermionic modes and integers for bosonic modes. At these certain frequencies, the gravity computation for correlator (\ref{eq:correlator}) matches precisely with CFT result trough Eq. (\ref{eq:correlatormatch}) up to a numerical normalization factor.

\subsection{Hidden conformal symmetry in $Q$-Picture}
\label{sec:scatteringonKSdSQpic}
In the previous subsections, we have shown the existence of hidden conformal symmetry on the $J$-picture with the background charged rotating black hole in Nariai limit. In this subsection, we will further study the scattering of massless scalar field on the background of charged rotating black hole solution in Nariai limit in $Q$-picture. On the other hand, we will consider charged scalar field on the background solution. For some black hole backgrounds, $Q$-picture has been explored well to exhibit the existence of the conformal symmetries \cite{Saktideformed2019,SaktiNucPhysB2020,ChenHuangPRD2010,ChenChen2011,GhezelbashSiahaanCQG2013,GhezelbashSiahaanGRG2014}. Nonetheless, the $Q$-picture is not well-defined for non-extremal Kerr-Sen black holes \cite{GhezelbashSiahaanCQG2013}. In this section, we will investigate the $Q$-picture when the cosmological constant exists. Since we will only consider $Q$-picture, for simplicity, we may assume that $p=0$, so that $k=0$.

The hidden conformal symmetries can be explored by assuming a massless charged scalar probe in the background of charged rotating black hole solution in Nariai limit. It is given by
\begin{equation}
	(\nabla_{\alpha}-igA_\alpha)(\nabla^{\alpha}-igA^\alpha) \Phi = 0\label{KG1Q},
\end{equation}
where $g$ is the electric charge of the scalar probe. In $Q$-picture, in addition to the modes of asymptotic energy and angular momentum, we need to add the charge $g$ as the eigenvalue of the operator $\partial_\chi$ that denotes the additional internal direction $\chi$ to four dimensions. Actually, $g$ has a natural geometrical interpretation as the radius of extra circle when the the black hole solution is considered to be embedded into 5D. In fact, this coordinate possesses similar $U(1)$ symmetry like the azimuthal coordinate. However, so far, we do not find that 4D Kerr-Sen(-dS) black hole can be uplifted to 5D solution. It is different with Kerr-Newman(-dS) black hole that can be uplifted to 5D \cite{Hartman2009}. Here, we will just first assume the fifth coordinate to reveal the conformal symmetries of the charged probe in $Q$-picture. The ansatz in $Q$-picture is given by
\begin{equation}
	\Phi(\hat{t}, \hat{r}, \theta, \hat{\phi},\chi) = \mathrm{e}^{- i \omega \hat{t} + i n \hat{\phi}+ig\chi} R(\hat{r}) S(\theta)\label{phi-expand1Q}.
\end{equation}
Beside the low-frequency assumption, we need to assume also that the probe's charge to be very small ($gq\ll 1 $). By plugging Eq. (\ref{phi-expand1Q}) into Eq. (\ref{KG1Q}) and assuming small charge and low frequency, in the near-horizon region, we can find the radial equation 
\begin{eqnarray}
	&& \partial_{\hat{r}} \left[(\hat{r}-r_c)(\hat{r}-r_+) \partial_{\hat{r}}\right]R(\hat{r}) + \frac{r_c - r_+}{\hat{r} - r_c} A  R(\hat{r}) = 0 ,  \nonumber\\
	&& +  \left[\frac{r_c - r_+}{\hat{r} - r_+} B + C \right] R(\hat{r})=0, \label{eq:nearregionradialeqEMDAQ}
\end{eqnarray} 
where
\begin{equation}
	B=-\frac{\left[ (r_+^2 -d^2 + a^2) \omega - a n \Xi -gq(r_+ +d) \right]^2}{\upsilon^2(r_c - r_+)^2}, \nonumber\
\end{equation}
\begin{equation}
		A= \frac{ \left[ (r_c^2 -d^2 + a^2) \omega - a n \Xi -gq(r_c +d) \right]^2}{\upsilon^2 (r_c - r_+)^2}, ~~~ C= \frac{-K_l}{\upsilon} . \label{CEMDAQ}\nonumber\
\end{equation}

In $Q$-picture, we take $n=0$. As the previous computation for $J$-picture, we can identify that
\begin{equation}
	n_L =- \frac{\upsilon (r_c +r_+)}{4\left[(r_c +d) (r_+ +d)  -a^2\right]},\nonumber\
\end{equation}
\begin{equation}
 n_R =-\frac{\upsilon (r_c - r_+)}{4\left[(r_c +d) (r_+ +d)  -a^2\right]}, \label{nJEMDAQ}
\end{equation}
\begin{equation}
	T_L = \frac{\upsilon(r_c^2 + r_+^2-2d^2+2a^2)}{4\pi g\left[(r_c +d) (r_+ +d)  -a^2\right]},\nonumber\
\end{equation}
\begin{equation}
	 T_R =\frac{\upsilon(r^2_c - r^2_+)}{4\pi g\left[(r_c +d) (r_+ +d)  -a^2\right]}. \label{eq:tempCFTKSdSQ}
\end{equation}
The temperatures (\ref{eq:tempCFTKSdSQ}) are irregular. We cannot check directly that $(r_c +d) (r_+ +d) -a^2=0$ because we do not have the explicit analytical form of $r_c,r_+$. However, we can easily check that the form of CFT temperatures is similar with the case when the cosmological constant vanishes. So, when $1/l^2 =0 $, one can obtain
\begin{equation}
(r_+ +d) (r_- +d)  -a^2=0, \label{eq:tempCFTKSdSQwithnocosmo}\
\end{equation}
where
\begin{equation}
r_\pm = M \pm \sqrt{M^2+d^2-a^2-q^2}.
\end{equation}
This is the reason that $Q$- picture is not well-defined for Kerr-Sen black hole. With the similar form of temperatures, we conclude also that $Q$-picture for Kerr-Sen-dS black hole in Nariai limit is not well-defined.

\section{Hidden Conformal Symmetry on Nariai limit for Kerr-Newman-de Sitter solution}\label{sec:KNdS}
In the previous sections, we have carried out the calculation of the dual CFT for charged rotating black hole in Nariai limit obtained from KSdS solution and also the hidden conformal symmetry of the scalar probe in the black hole's background. In this section, we will study the scattering of scalar field on Nariai limit for KNdS solution both in $J$- and $Q$-pictures. Yet, we will revisit the computation of the entropy from CFT for KNdS solution in Nariai limit. The KNdS black hole solution is given by the following metric \cite{Hartman2009,Sinamuli2016}
\begin{eqnarray}
	ds^2 = - \frac{\Delta}{\varrho^2} X^2 +  \frac{\varrho ^2}{\Delta}d\hat{r}^2 + \frac{\varrho ^2}{\Delta_\theta} d\theta^2   + \frac{\Delta_\theta\sin^2\theta}{\varrho ^2}Y^2,\label{eq:KNdSmetric} \
\end{eqnarray}
where
\begin{equation}
	X=d\hat{t} - \frac{a \sin^2\theta}{\Xi} d\hat{\phi},~~~ Y=ad\hat{t} - \frac{(\hat{r}^2+a^2)}{\Xi}  d\hat{\phi}, \nonumber\
\end{equation}
%\vspace{-0.7cm}
\begin{equation}
	\Delta_\theta = 1+\frac{a^2}{l^2}\cos^2\theta, ~~\Xi =1+\frac{a^2}{l^2}, ~~~ \varrho^2 =\hat{r}^2 + a^2\cos^2\theta,\nonumber\\
\end{equation}
%\vspace{-0.7cm}
\begin{equation}
	\Delta = (\hat{r}^2+a^2)\left(1-\frac{\hat{r}^2}{l^2} \right)-2m\hat{r}+p^2 +q^2.\
\end{equation}
The parameters $a$, $m$, $p$, $q$, and $l$  are spin, mass, magnetic charge, electric charge, and de Sitter radius, respectively. The electromagnetic potential and its dual are given by
%\begin{eqnarray}
%	\textbf{A} = \frac{q\hat{r}-pa\cos\theta}{\varrho^2}X+\frac{P(\cos\theta\mp\hat{\sigma)}}{\Xi}d\hat{\phi}, \label{eq:AKNdS}
%\end{eqnarray}
%while its dual is given by
%\begin{eqnarray}
%	\textbf{B} = \frac{p\hat{r}+qa\cos\theta}{\varrho^2}X-\frac{P(\cos\theta\mp\hat{\sigma)}}{\Xi}d\hat{\phi}, \label{eq:BKNdS}
%\end{eqnarray}
\begin{eqnarray}
	\textbf{A} =- \frac{q\hat{r}}{\varrho^2}X-\frac{p\cos\theta}{\varrho^2}Y,~~~\textbf{B} =- \frac{p\hat{r}}{\varrho^2}X+\frac{q\cos\theta}{\varrho^2}Y, \label{eq:ABKNdSshift}
\end{eqnarray}

At the event horizon, the thermodynamic quantities are given by \cite{Hartman2009,Sinamuli2016}
\begin{equation}
	M= \frac{m}{\Xi}, ~~~ J = \frac{ma}{\Xi}, ~~~ Q= \frac{q}{\Xi}, ~~~ P= \frac{p}{\Xi},  \label{eq:MJQPKNdS}
\end{equation}
%\vspace{-0.6cm}
\begin{equation}
	T_H = \frac{r_+(l^2-2r_+^2-a^2 )-ml^2}{2\pi (r_+^2 +a^2)l^2}, \label{eq:THKNdS}
\end{equation}
%\vspace{-0.6cm}
\begin{equation}
	S_{BH}=\frac{\pi}{\Xi}(r_+^2 +a^2), \label{eq:SBHKNdS}
\end{equation}
%\vspace{-0.6cm}
\begin{equation}
	\Omega_H = \frac{a\Xi}{r_+^2 +a^2}, ~~
	\Phi_H = \frac{qr_+}{r_+^2+a^2},~~	\Psi_H = \frac{pr_+}{r_+^2 +a^2}, \label{eq:PhiPsiKNdS}
\end{equation}
%\vspace{-0.6cm}
\begin{equation}
	V= \frac{4}{3}r_+ S_{BH}, ~~~\mathcal{P}=-\frac{3}{8\pi l^2}. \label{eq:VKNdS}
\end{equation}
where those are physical mass, angular momentum, electric charge, magnetic charge, Hawking temperature, Bekenstein-Hawking entropy, angular velocity, electric potential, magnetic potential, thermodynamic volume and pressure. These thermodynamic quantities also satisfy the relation (\ref{eq:thermoKSdS}). Furthermore, one can find the similar relation for thermodynamics on the cosmological horizon with the following quantities \cite{DehganiPRD2002,GhezelbashMannPRD2005}
\begin{equation}
	M_c= -\frac{m}{\Xi}, ~~~ J_c = -\frac{ma}{\Xi}, ~~~ Q_c= -\frac{q}{\Xi}, ~~~ P_c= -\frac{p}{\Xi},  \label{eq:MJQPKNdSCos}
\end{equation}
%\vspace{-0.6cm}
\begin{equation}
	T_c = \frac{r_c(2r_c^2+a^2-l^2 )+ml^2}{2\pi (r_c^2 +a^2)l^2}, \label{eq:THKNdSCos}
\end{equation}
%\vspace{-0.6cm}
\begin{equation}
	S_{c}=\frac{\pi}{\Xi}(r_c^2 +a^2), \label{eq:SBHKNdSCos}
\end{equation}
%\vspace{-0.6cm}
\begin{equation}
	\Omega_c = \frac{a\Xi}{r_c^2 +a^2}, ~~
	\Phi_c = \frac{qr_c}{r_c^2+a^2},~~	\Psi_c = \frac{pr_c}{r_c^2 +a^2}, \label{eq:PhiPsiKNdSCos}
\end{equation}
%\vspace{-0.6cm}
\begin{equation}
	V_c= \frac{4}{3}r_c S_c, ~~~\mathcal{P}_c=\frac{3}{8\pi l^2}. \label{eq:VKNdSCos}
\end{equation}
These thermodynamic quantities are obtained also by considering the event horizon of the black hole as the boundary. Likewise the KSdS solution, the cosmological constant can be assumed as a dynamical quantity.

\subsection{Nariai limit on Kerr-Newman-de Sitter black hole revisited}
On the Nariai limit, likewise the KSdS solution, we can find the geometry which is fiber over AdS$_2$. Using the similar transformations (\ref{eq:Nariaipar}) and (\ref{eq:nearhorizontrans}) on the spacetime metric (\ref{eq:KSdSmetric}), we can obtain
\begin{eqnarray}
	ds^2 &=& \Gamma(\theta)\left(-r(r-\epsilon) dt^2 + \frac{dr^2}{r(r-\epsilon)} + \alpha(\theta) d\theta ^2 \right)  \nonumber\\
	&& +\gamma(\theta) \left(d\phi +e r dt\right)^2, \label{eq:rotatingNariaimetricEMDA}\
\end{eqnarray}
where the metric functions are given by
\begin{equation}
	\Gamma(\theta)= \frac{\varrho_+ ^2}{\upsilon} , ~~\alpha(\theta) = \frac{\upsilon}{\Delta_\theta}, ~~\gamma(\theta)=\frac{r_0^4 \Delta_\theta \sin^2\theta}{\varrho_+^2 \Xi^2},\nonumber\
\end{equation}
\begin{equation}
	\varrho_+^2 = r_+^2+ a^2\cos^2\theta, ~~
	e = \frac{2ar_+\Xi}{r_0^2 \upsilon}. \label{eq:constant_eEMDA}
\end{equation}
and now we have $r_0^2=r_+^2+a^2$ and $\upsilon = 1-(6r_+^2+a^2)/l^2$. When the spin vanishes, the spacetime becomes AdS$_2\times S^2$. 

Even we have different slice in the asymptotic region, by using similar analysis of ASG, we can obtain the exact central charge as found in \cite{Anninos2010}. It is precisely given by
\begin{equation}
c_L=\frac{12ar_+}{1-\frac{6r_+^2+a^2}{l^2}}.\label{eq:cLKNdScos}\
\end{equation}
The main different of this central charge with that of the solution in Nariai limit obtained from KSdS solution is the presence of the dilaton and axion charges. The temperature can be calculated directly as the previous result for Nariai limit on KNdS solution. We can find that 
\begin{eqnarray}
T_L = \frac{\upsilon(r_+^2+a^2)}{4\pi ar_+ \Xi},~~ T_R=0. \label{eq:TLKNdScos}\
\end{eqnarray}

We have obtained the corresponding central charge and the CFT temperature for the solution in Nariai limit obtained from KNdS black hole. On the cosmological horizon, by using Cardy formula, we find the following entropy for the charged rotating black hole from KNdS black hole in Nariai limit,
\begin{equation}
S_{CFT}=\frac{\pi}{\Xi}(r_+^2+a^2)=S_{BH}=S_c.\label{eq:entropyCFTNariaiKNdS}\
\end{equation}
So, in this case, the total entropy will be $2S_c$.

\subsection{Hidden conformal symmetry in $J$-picture}
For the charged rotating solution in Nariai limit obtained from KNdS black hole, we also assume the massless neutral scalar probe. We also consider the low-frequency limit for the scalar field to exhibit the conformal symmetry. For the small frequency, we have $ \omega M \ll 1 $, $ \omega a \ll 1, \, \omega q \ll 1,  $ and $\omega p \ll 1$. The radial equation is then given by 
\begin{eqnarray}
&& \partial_{\hat{r}} \left[(\hat{r}-r_c)(\hat{r}-r_+) \partial_{\hat{r}}\right]R(\hat{r}) + \frac{r_c - r_+}{\hat{r} - r_c} A  R(\hat{r}) = 0 , \nonumber\\
&& +  \left[\frac{r_c - r_+}{\hat{r} - r_+} B + C \right] R(\hat{r})=0, \label{eq:nearregionradialeq}
\end{eqnarray} 
where
\begin{equation}
A= \frac{ \left[ (r_c^2+ a^2) \omega - a n \Xi \right]^2}{\upsilon^2 (r_c - r_+)^2}, \nonumber\
\end{equation}
\begin{equation}
B=-\frac{\left[ (r_+^2+ a^2) \omega - a n \Xi \right]^2}{\upsilon^2(r_c - r_+)^2}, ~~~ C= \frac{-K_l}{\upsilon} . \label{C}\nonumber\
\end{equation}
The investigation of hidden conformal symmetry for KNdS solution is also done in Ref.~\cite{SiahaanEPJC2022}, however, the author does not consider Nariai limit. The conformal symmetry can be revealed using the conformal coordinates transformations (\ref{conformalcoord}) which similarly result in $ SL(2,R)\times SL(2,R) $ isometry for the set of operators (\ref{vec1}) and (\ref{vvec1}). Using the similar computation, we can calculate the temperatures.
In this case, we obtain
\begin{equation}
	n_L = -\frac{\upsilon}{2(r_c + r_+)},~~~ n_R =0, \label{nJ}
\end{equation}
\begin{equation}
	T_L = \frac{\upsilon(r_c^2 + r_+^2+2a^2)}{4\pi a(r_c + r_+)\Xi},~~~ T_R =\frac{\upsilon(r_c - r_+)}{4\pi a\Xi}. \label{eq:tempCFTKNdS}
\end{equation}
Note that those quantities are different to those of the Nariai solution obtained from KSdS solution since there exist dilaton and axion charges. When we take $\varepsilon\rightarrow0$, the temperatures (\ref{eq:tempCFTKNdS}) reduce to (\ref{eq:TLKNdScos}).
 
After computing the CFT temperatures (\ref{eq:tempCFTKNdS}), we will compute the conjugate charges $E_L,E_R$ for this solution. These conjugate charges can be obtained from the entropy via Eq. (\ref{eq:variationCFTentropy}) by considering the first law of thermodynamics for the charged rotating black hole solution (\ref{eq:thermoKSdS}) with the quantities (\ref{eq:MJQPKNdSCos})-(\ref{eq:VKNdSCos}). For the neutral scalar field, we have $\delta Q= \delta P=\delta\mathcal{P}=0$. Using again the identification $\delta M$ as $\omega$ and $\delta J$ as $n$ yields to the identification of $\delta E _{L,R}$ as $\omega_{L,R}$.  We obtain the following left and right frequencies,
\begin{equation}
\omega _{L} = \frac{r_c^2 + r_+^2 +2a^2}{2a\Xi}\omega, ~~~\omega _{R}=\omega _{L}-n.  \label{eq:omegaCFTgauged} \
\end{equation}
These are the left and right frequencies of 2D CFT for the charged rotating solution for KNdS solution in Nariai limit.
	
Regarding the absorption cross-section and real-time correlator, we need to employ the radial equation (\ref{eq:nearregionradialeq}). Similarly with KNdS case, the approximation of $\Delta$ in the asymptotic region will break down, except we consider the Nariai limit which is identical to near-extremal limit. We also consider the scalar probe with frequencies around the superradiant bound (\ref{eq:superbound}). With the similar lengthy computation, we find the absorption cross-section of the scalar fields as given in Eq. (\ref{eq:Pabsgauged}) which is precisely similar to (\ref{eq:PabsCFT}) for the charged rotating solution in Nariai limit obtained from KNdS solution. Once again, the agreement between (\ref{eq:Pabsgauged}) and (\ref{eq:PabsCFT}) for this solution can be shown when we choose proper left and right frequencies $\omega_L,\omega_R$ as given in Eq. (\ref{eq:omegaCFTgauged}). In the superradiant bound, we also find the form of temperatures (\ref{eq:temperatureCFTnariai}) and the frequencies (\ref{eq:omegasuper}) with the conformal weights (\ref{eq:conformalweightCFT}). Yet, we have to note again that in the solution coming from Einstein-Maxwell theory, there is no contribution from $d,k$ on $r_+$. We can also compute the real-time correlator. This is given by Eq. (\ref{eq:retardedcorr1}). This further exhibits that the Kerr/CFT correspondence for KNdS black hole in Nariai limit is valid likewise the solution from KSdS black hole.

\subsection{Hidden conformal symmetry in $Q$-picture}

In exploring the hidden conformal symmetries on the wave equation, again, we assume a massless charged scalar probe in the background (\ref{eq:KNdSmetric}) in Nariai limit as given by Eq. (\ref{KG1Q}). By plugging Eq. (\ref{phi-expand1Q}) into Eq. (\ref{KG1Q}) and assuming small charge and low frequency, in the near-horizon region of (\ref{eq:KNdSmetric}) in Nariai limit, we can find the radial equation 
\begin{eqnarray}
	&& \partial_{\hat{r}} \left[(\hat{r}-r_c)(\hat{r}-r_+) \partial_{\hat{r}}\right]R(\hat{r}) + \frac{r_c - r_+}{\hat{r} - r_c} A  R(\hat{r}) = 0 ,  \nonumber\\
	&& +  \left[\frac{r_c - r_+}{\hat{r} - r_+} B + C \right] R(\hat{r})=0, \label{eq:nearregionradialeqQ}
\end{eqnarray} 
where
\begin{equation}
	B=-\frac{\left[ (r_+^2 + a^2) \omega - a n \Xi -gq r_+ \right]^2}{\upsilon^2(r_c - r_+)^2}, \nonumber\
\end{equation}
\begin{equation}
	A= \frac{ \left[ (r_c^2 + a^2) \omega - a n \Xi -gqr_c \right]^2}{\upsilon^2 (r_c - r_+)^2}, ~~~ C= \frac{-K_l}{\upsilon} . \label{ABCQ}\nonumber\
\end{equation}
For this charged rotating solution obtained from KNdS black hole in Nariai limit, in $Q$-picture, we can identify that
\begin{equation}
	n_L =- \frac{\upsilon (r_c +r_+)}{4\left(r_c r_+ -a^2\right)},~
	n_R =-\frac{\upsilon (r_c - r_+)}{4\left(r_c r_+ -a^2\right)}, \label{nQ}
\end{equation}
\begin{equation}
	T_L = \frac{\upsilon(r_c^2 + r_+^2+2a^2)}{4\pi g\left(r_c r_+  -a^2\right)}, ~
	T_R =\frac{\upsilon(r^2_c - r^2_+)}{4\pi g\left(r_c r_+  -a^2\right)}. \label{eq:tempCFTQ}
\end{equation}
These CFT temperatures are regular, unlike to that of KSdS solution in Nariai limit.

Since there exists electric charge of the scalar probe, in order to satisfy the entropy relation (\ref{eq:variationCFTentropy}), we need the conjugate of the electric charge, namely the chemical potential ($\mu_{L,R}$), so that
\begin{equation}
E_{L,R} = \hat{\omega}_{L,R}= \omega_{L,R}-\mu_{L,R} q_{L,R}.\label{eq:CFTfrequenciesKNdS}\
\end{equation}
So, in $Q$-picture, the charged scalar probe on the charged rotating solution obtained from KNdS black hole in Nariai limit is related to the following CFT frequencies,
\begin{eqnarray}
\omega_{L}&=&\frac{(r_c +r_+)(r_c^2 +r_+^2+2a^2)}{2g(r_c r_+ -a^2)}\omega, \nonumber\\
\mu_L &=& \frac{r_c^2 +r_+^2+2a^2}{2(r_c r_+ -a^2)}, ~~~ q_L=q, \nonumber\\
\omega_{R}&=&\omega_L -\frac{2a(r_c +r_+)}{2g(r_c r_+ -a^2)}n, \nonumber\\
\mu_R &=& \frac{(r_c+r_+)^2}{2(r_c r_+ -a^2)}, ~~~ q_R=q. \
\end{eqnarray}
One can apply the frequencies (\ref{eq:CFTfrequenciesKNdS}), temperatures (\ref{eq:tempCFTQ}), and conformal weights (\ref{eq:conformalweightCFT}) in order to compute the absorption cross-section and real-time correlator. Note that to study the superradiant bound in $Q$-picture, we have assume the frequency near the superradiant bound $\omega = n\Omega_H +g\Phi_H+\hat{\omega}\frac{\lambda}{r_0}$.

\section{Summary}\label{sec:summary}
In this work, we have shown the horizon solutions and extremal limits in Fig. \ref{figFin}. The Nariai limit could be achieved in some values of parameters.
We then have carried out the calculation of the entropy for KSdS solutions in Nariai limit parameterized by a constant $\varepsilon$. When $\varepsilon\rightarrow 0$, the cosmological horizon and event horizon coincide. This solution in Nariai limit can be written in the metric form with AdS$_2$ structure. Hence, we have shown that, in the Nariai limit, we could also portray the solution as a fiber over AdS$_2$, instead of a fiber over dS$_2$ as exhibited in Ref.~\cite{Anninos2010}. We have computed the corresponding central charge and CFT temperature where the right-moving temperature is proportional to $\varepsilon$ denoting that it will vanish when $\varepsilon\rightarrow 0$. It is found that by employing Cardy entropy formula, the Bekenstein-Hawking entropy on the cosmological constant is reproduced. It denotes that the charged rotating solutions from KSdS black hole in Nariai limit is holographically dual with 2D CFT.

To further support the CFT dual on KSdS black hole in Nariai limit, we have investigated the neutral ($J$-picture) and charged ($Q$-picture) massless scalar probes on that background. Similarly with generic rotating black hole, we could exhibit the conformal symmetries on the radial wave equation. With the appropriate locally conformal coordinate transformations, it has been shown that the radial equation possesses $SL(2,R)\times SL(2,R)$ isometry. This is similar with AdS$_3$ space. The periodic identification of azimuthal coordinate portrays the spontaneous symmetry breaking from $SL(2,R)\times SL(2,R)$ to $U(1)\times U(1)$ by the left- and right-moving temperatures. In $J$-picture, the temperatures produced on this symmetry breaking are precisely similar with the temperatures that are obtained when the conformal symmetry appears directly on the spacetime metric. Hence, if we employ the central charges with the given temperatures, we can again reproduce the Bekenstein-Hawking entropy on the cosmological horizon. Moreover, we also have computed the absorption cross-section and the real-time correlator that correspond with the KSdS solution in Nariai limit. However, in $Q$-picture, we coud not find a well-defined CFT description because the temperatures are irregular.

We have also extended the calculation to the KNdS black hole in Nariai limit  both in $J$- and $Q$-pictures. The CFT description in both pictures are well-defined. The results from gravity calculation are also exactly in agreement with the result from 2D CFT. So, this calculation is another proof that the black hole solution in Nariai limit is holographically dual to 2D CFT.

\appendix
\section{Entropy on the Event Horizon for Extremal KSdS solution}
\label{app:entropyKSdS}
We will briefly derive the entropy of extremal KSdS solution in this section. The entropy of the gauged dyonic Kerr-Sen black hole solution or KSAdS solution has been calculated previously in Ref.~\cite{SaktidyonicKerrSen}. For the asymptotically de Sitter solution, we just need to follow the computation given in Ref.~\cite{SaktidyonicKerrSen} where therein, we have argued on the central charge for KSdS solution by taking $l^2 \rightarrow -l^2$. It has been noted in Ref.~\cite{SaktidyonicKerrSen} that in the extremal case of KSAdS solution, the mass can have more than two branches as it should be similar to KSdS solution. Let start directly from the near-horizon extremal form of KSdS solution,
\begin{eqnarray}
	ds^2 &=& \Gamma(\theta)\left(-r^2 dt^2 + \frac{dr^2}{r^2} + \alpha(\theta) d\theta ^2 \right) \nonumber\\
	&& +\gamma(\theta) \left(d\phi +e r dt\right)^2, \label{eq:extremalmetricdKSdS}\
\end{eqnarray}
where the metric functions are given by
\begin{equation}
\Gamma(\theta)= \frac{\varrho_+ ^2}{\upsilon} , ~~~\alpha(\theta) = \frac{\upsilon}{\Delta_\theta}, ~~~\gamma(\theta)=\frac{r_0^4 \Delta_\theta \sin^2\theta}{\varrho_+^2 \Xi^2}, \nonumber\
\end{equation}
\begin{equation}
\varrho_+^2 = r_+^2 -d^2 -k^2 + a^2\cos^2\theta, ~~~
e = \frac{2ar_+\Xi}{r_0^2 \upsilon}. \label{eq:constant_eKSdS}\nonumber\
\end{equation}
\begin{equation}
\upsilon = 1 - \frac{6r_+^2-2d^2-2k^2+a^2}{l^2}.\label{eq:upsilonKSdS}
\end{equation}
In this case, we can precisely derive the corresponding central charge as
\begin{eqnarray}
	c_L&=&\frac{3e}{\Xi}\int^\pi_0d\theta\sqrt{\Gamma(\theta)\alpha(\theta)\gamma(\theta)}\nonumber\\
	&=&\frac{12ar_+}{1-\frac{6r_+^2 -d^2-k^2+a^2}{l^2}}.\label{eq:cLKSdS}\
\end{eqnarray}
Note that although this central charge looks identical with that of in Nariai limit, both central charges are differnt because we are considering different limit on the horizons. When the magnetic charge vanishes, one can obtain the central charge that corresponds to non-dyonic KSAdS solution. Furthermore, when the cosmological constant vanishes, the central charge recovers the central charge of Kerr-Sen black hole \cite{Ghezelbash2009,GhezelbashSiahaanCQG2013}. For the CFT temperature, we just need to follow Ref.~\cite{SaktidyonicKerrSen}. For KSdS solution, we can find
\begin{eqnarray}
	T_L = - \frac{\partial T_H/\partial r_+}{\partial \Omega_H / \partial r_+}\bigg|_{ex} =\frac{\upsilon(r_+^2-d^2-k^2+a^2)}{4\pi ar_+ \Xi}. \label{eq:TLKSdS}\
\end{eqnarray}
By implementing Cardy entropy formula and inserting (\ref{eq:cLKSdS}) and (\ref{eq:TLKSdS}) into it, we successfully derive
\begin{equation}
	S_{CFT}=\frac{\pi}{\Xi}(r_+^2-d^2-k^2+a^2).\label{eq:entropyCFTKSdS}\
\end{equation}
This is precisely the Bekenstein-Hawking entropy of the KSdS black hole on the event horizon. This result shows that we can change $l^2$ on those of KSAdS solution into $-l^2$ by assuming analytical continuation in order to find the result on those of KSdS solution. This is identical with the KNAdS(dS) case \cite{Hartman2009}.

\section{Virasoro algebra}
\label{app:Virasoro}
The Dirac bracket of two classical conserved charges are given by
\begin{eqnarray}
	\left\{ Q_{\zeta_m},Q_{\zeta_n} \right\} = Q_{[\zeta_m,\zeta_n]} + K[\zeta_m,\zeta_n], \label{eq:Diracbracket}
\end{eqnarray}
where $ K[\xi,\zeta]$ is the central term. Using the quantum version of charge $Q_\xi$ (\ref{eq:quantumQL}), from Eq. (\ref{eq:Diracbracket}) we can find
\begin{eqnarray}
	[L_m,L_n]&=&i\{Q_{\zeta_m},Q_{\zeta_n}\}\nonumber\\
	&=& i\left(Q_{[\zeta_m,\zeta_n]}+K[\zeta_m,\zeta_n]\right)\\
	&=& (m-n)L_{m+n}-2mx\delta_{m+n}+i K[\zeta_m,\zeta_n].\nonumber\
\end{eqnarray}
In order to find the Virasoro algebra, we need to find the form of term $K[\zeta_m,\zeta_n]$. By comparing above equation with Virasoro algebra,
\begin{equation}
	\left[L_{m} , L_{n}  \right] = ({m} - {n}) L_{m+n} + \frac{c_L}{12}m (m ^2-1)\delta_{m+n , 0}, 
\end{equation}
we obtain 
\begin{equation}
	K[\zeta_m,\zeta_n]=-i\frac{c_L}{12}m\left(m^2-1+\frac{24x}{c_L}\right)\delta_{m+n,0}.\label{eq:Ktermnew}
\end{equation}
So, the central charge is determined by the coefficient $m^3$ in term $K[\zeta_m,\zeta_n]$. The term linear in $m$ is not important because $x$ is a free parameter to scale the last term in the bracket of Eq. (\ref{eq:Ktermnew}). 

%\end{verbatim}
%will produce an appendix heading that says ``APPENDIX A'' and
%\begin{verbatim}
%\appendix
%\section{Background}
%\end{verbatim}
%will produce an appendix heading that says ``APPENDIX A: BACKGROUND''
%(note that the colon is set automatically).
%
%If there is only one appendix, then the letter ``A'' should not
%appear. This is suppressed by using the star version of the appendix
%command (\verb+\appendix*+ in the place of \verb+\appendix+).
%
%\section{A little more on appendixes}
%
%Observe that this appendix was started by using
%\begin{verbatim}
%\section{A little more on appendixes}
%\end{verbatim}
%
%Note the equation number in an appendix:
%\begin{equation}
%E=mc^2.
%\end{equation}
%
%\subsection{\label{app:subsec}A subsection in an appendix}
%
%You can use a subsection or subsubsection in an appendix. Note the
%numbering: we are now in Appendix~\ref{app:subsec}.
%
%Note the equation numbers in this appendix, produced with the
%subequations environment:
%\begin{subequations}
%\begin{eqnarray}
%E&=&mc, \label{appa}
%\\
%E&=&mc^2, \label{appb}
%\\
%E&\agt& mc^3. \label{appc}
%\end{eqnarray}
%\end{subequations}
%They turn out to be Eqs.~(\ref{appa}), (\ref{appb}), and (\ref{appc}).

% The \nocite command causes all entries in a bibliography to be printed out
% whether or not they are actually referenced in the text. This is appropriate
% for the sample file to show the different styles of references, but authors
% most likely will not want to use it.
%\nocite{*}
\vspace*{8mm}

\noindent {\large{\bf Acknowledgments}}
\vspace*{4mm}

The authors thank the reviewer for the fruitful suggestions on this manuscript. This work is supported by the Second Century Fund (C2F), Chulalongkorn University, Thailand. P. B. is supported in part by National Research Council of Thailand~(NRCT) and Chulalongkorn University under Grant N42A660500. This research has received funding support from the NSRF via the Program Management Unit for Human Resources \& Institutional Development, Research and Innovation~[grant number B39G660025].  \newline
\vspace*{1mm}

\bibliography{apssamp}% Produces the bibliography via BibTeX.

\end{document}